%% file: arxiv.tex
\documentclass[aip,jcp,amsmath,amssymb,reprint,floatfix]{revtex4-1}
\input{header.tex}
\begin{document}

\title{Ab Initio Effective One-Electron Potential Operators. I.
Applications for Charge-Transfer Energy in Effective Fragment Potentials}

\author{Bartosz B{\l}asiak}
\email[]{blasiak.bartosz@gmail.com}
\homepage[]{https://www.polonez.pwr.edu.pl}

\author{Joanna D. Bednarska}
\author{Marta Cho{\l}uj} 
\author{Wojciech Bartkowiak}

\affiliation{Department of Physical and Quantum Chemistry, Faculty of Chemistry, 
Wroc{\l}aw University of Science and Technology, 
Wybrze{\.z}e Wyspia{\'n}skiego 27, Wroc{\l}aw 50-370, Poland}

\date{\today}

\begin{abstract}
The concept of the effective one\hyp{}electron potentials (OEP) has been useful for many decades
in efficient description of electronic structure of chemical systems, especially extended
molecular aggregates such as interacting molecules in condensed phases.
Here, a general method for effective OEP\hyp{}based elimination of electron repulsion integrals (ERI),
that is tuned
towards the fragment\hyp{}based calculation methodologies
such as the second generation of the effective fragment potentials (EFP2) method,
is presented.
Two general types of the OEP operator matrix elements
are distinguished and treated either via the distributed multipole expansion
or the extended density fitting schemes developed in this work. 
The OEP technique is then applied
to address the problem of using incomplete EFP2 settings in many
applications in interaction energy and 
molecular dynamics simulations 
due to relatively high computational cost of evaluating the
charge transfer (CT) effects as compared to other effects.
The alternative OEP\hyp{}based CT energy model is proposed
in the context of the
intermolecular perturbation theory with Hartree\hyp{}Fock non\hyp{}interacting gas\hyp{}phase 
reference wavefunctions, compatible with the EFP2 formulation.
It is found that the computational cost can be reduced
up to 20 times as compared to the CT energy method within the EFP2 scheme
without compromising the accuracy for a wide range of weakly interacting 
neutral molecular complexes. 
Therefore, it is believed that
the proposed model can be used within 
the EFP2 framework, making the CT energy term no longer the bottleneck
in EFP2\hyp{}based simulations of complex systems.
\end{abstract}


\maketitle

%\tableofcontents

\section{\label{s:1.introduction}Introduction}

Charge transfer (CT) between molecules occurs when the net electronic populations 
of interacting molecules change which leads to an additional stabilization 
of a molecular aggregate.\cite{Otto.Ladik.IJQC.1980,Stone.TheTheoryOfIntermolecularForces.1996}
Although CT processes can be relatively easily scrutinized based
on the total amount of electronic charge transferred between interacting species,\cite{Otto.Ladik.IJQC.1980}
evaluation of its contribution to the intermolecular interaction
potential\cite{Stone.TheTheoryOfIntermolecularForces.1996} is far from trivial due to 
its notably complex quantum mechanical (QM) origins 
even at the Hartree\hyp{}Fock\cite{Roothaan.RevModPhys.1951} (HF)
approximation level,\cite{Jensen.JCP.2001,Otto.Ladik.ChemPhys.1975}
and cannot be realized in terms of any classical nor semi\hyp{}classical approach.
In fact, CT does not naturally emerge in the symmetry 
adapted perturbation theory\cite{Jeziorski.Moszynski.Szalewicz.ChemRev.1994} (SAPT)
due to technical aspects related with the use of finite basis set expansions
in modern quantum chemistry calculations and avoiding the basis set superimposition error (BSSE). 
Stone and Misquitta showed that CT energy
can be extracted from SAPT calculations by comparing the induction energies
of fictional systems in which basis functions are centered either on the monomers only, or the entire 
interacting complex.\cite{Stone.Misquitta.CPL.2009}
CT energy was formulated by Murrell et al.~\cite{Murrell.Randic.Williams.Longuet-Higgins.ProcRSocLondA.1965} 
in their perturbation theory in the region of small wavefunction overlap up to second order.
However, all these theories are computationally too expensive to apply for efficient
calculation of intermolecular forces in molecular dynamics because
they involve calculation of the electron repulsion integrals (ERI's)
and their four\hyp{}index transformation to molecular orbital (MO) basis.\cite{Otto.Ladik.ChemPhys.1975}

The apparent difficulty in theoretically characterizing the CT energy
in terms of the interacting molecular fragments
is indeed a challenge in the development of modern force fields 
or \emph{ab initio} fragmentation methods\cite{Gordon.Fedorov.Pruitt.Slipchenko.ChemRev.2012}
for modeling structure
and dynamics in condensed phases.\cite{Demerdash.Mao.Liu.Head-Gordon.Head-Gordon.JCP.2017}
This needs to be contrasted with the Coulombic electrostatics,
non\hyp{}Coulombic repulsion (i.e., due to Pauli exclusion principle), 
dispersion and induction for instance, 
which to a certain extent
can be well described by
relatively simple and computationally
inexpensive to evaluate mathematical models, i.e.,
the distributed multipole moments of charge densities,\cite{Sokalski.Poirier.CPL.1983,
Etchebest.Lavery.Pullman.TheorChimActa.1982,Stone.JCTC.2005}
the van der Waals repulsive and attractive potentials, 
as well as the polarizability models. 
Due to this reason, the CT effects 
are not explicitly included in most of molecular mechanics force field 
developed up to date.\cite{Demerdash.Yap.Head-Gordon.AnnuRevPhysChem.2014}
There is only a few force fields which
explicitly
incorporate the CT effects in the condensed\hyp{}phase simulations
in \emph{ab initio} manner,\cite{Xu.Guidez.Bertoni.Gordon.JCP.2018}
such as the 
second generation of the Effective Fragment Potential 
(EFP2) method\cite{Gordon.Smith.Xu.Slipchenko.AnnuRevPhysChem.2013,
   Nguyen.Pachter.Day.JCP.2014,
   Day.Jensen.Gordon.Webb.Stevens.Krauss.Garmer.Basch.Cohen.JCP.1996,
   Sattasathuchana.Xu.Gordon.JPCA.2019,
   Kuroki.Mori.ChemLett.2016,
   Ghosh.Cho.Choi.JPCB.2014,
   Jensen.JCP.2001,Li.Gordon.Jensen.JCP.2006,Xu.Gordon.JCP.2013}
or Sum of Interactions Between Fragments 
Ab initio Computed (SIBFA) method\cite{Gresh.Claverie.Pullman.TCA.1984,Piquemal.Chevreau.Gresh.JCTC.2007},
apart from performing full QM electronic structure simulations
or non\hyp{}force\hyp{}field\hyp{}based 
fragmentation techniques\cite{Leverentz.Maerzke.Keasler.Siepmann.Truhlar.PCCP.2012,
Dahlke.Truhlar.JCTC.2008}.
EFP2, that is one of the most commonly used force fields of this kind,
was derived on the grounds of the first\hyp{}principles %perturbation theory
at the HF level\cite{Jensen.JCP.2001,Jensen.Gordon.MolPhys.1996,Jensen.Gordon.JCP.1998,
Li.Netzloff.Gordon.JCP.2006,Li.Gordon.Jensen.JCP.2006,Xu.Gordon.JCP.2013} 
and including the intermolecular dispersion effects
by the response theory.\cite{Adamovic.Gordon.MolPhys.2005,Xu.Zahariev.Gordon.JCTC.2014}
That is to say, the total intermolecular 
interaction potential
%which can also be perceived as an approximation
%to SAPT at HF level with added intermolecular dispersion correction, 
is approximated as 
\begin{equation}\label{e:u-efp2}
 E^{\rm EFP2} \approx E^{\rm Coul} + E^{\rm Ex-Rep} + E^{\rm Ind} + E^{\rm Disp} + E^{\rm CT} \;,
\end{equation}
where $E^{\rm Coul}$ is the Coulombic interaction energy of the unperturbed charge\hyp{}density
distributions of the monomers, treated by the distributed multipole approximation
with damping to account for the charge\hyp{}penetration effects,\cite{Slipchenko.Gordon.JCC.2007}
$E^{\rm Ex-Rep}$ is the exchange\hyp{}repulsion energy originating from the Pauli exclusion
principle,\cite{Jensen.Gordon.MolPhys.1996,Jensen.Gordon.JCP.1998} 
that could be roughly understood as the Heitler\hyp{}London interaction
energy\cite{Chalasinski.Gutowski.MolPhys.1985} with $E^{\rm Coul}$ subtracted, $E^{\rm Ind}$ and $E^{\rm Disp}$
are the induction and dispersion energies obtained from the distributed polarizability
approximation,\cite{Li.Netzloff.Gordon.JCP.2006,Adamovic.Gordon.MolPhys.2005,Xu.Zahariev.Gordon.JCTC.2014} 
and finally $E^{\rm CT}$ is the CT energy,\cite{Li.Gordon.Jensen.JCP.2006,Xu.Gordon.JCP.2013} 
the focus of this work.

Despite the considerable success of the EFP2 theory
in accurately modeling the extended molecular systems 
like solutions\cite{Sattasathuchana.Xu.Gordon.JPCA.2019,
   Blasiak.Londergan.Webb.Cho.ACR.2017,
   Kuroki.Mori.ChemLett.2016,
   Ghosh.Cho.Choi.JPCB.2014}
and recently even biomolecules\cite{Blasiak.Ritchie.Webb.Cho.PCCP.2016,
Xu.Blasiak.Cho.Layfield.Londergan.JPCL.2018,
Gurunathan.Acharya.Ghosh.Kosenkov.Kaliman.Shao.Krylov.Slipchenko.JPCB.2016,
Ghosh.Kosenkov.Vanovschi.Williams.Herbert.Gordon.Schmidt.Slipchenko.Krylov.JPCA.2010}
with the level of accuracy
reaching in many cases\cite{Gordon.Smith.Xu.Slipchenko.AnnuRevPhysChem.2013} 
the M{\o}ller\hyp{}Plesset perturbation theory\cite{Moller.Plesset.PhysRev.1934},
evaluation of the CT energy in EFP2 model is still relatively costly for
typical uses in the
molecular dynamics simulations. It has been reported that 
the implementation of the EFP2 CT energy and gradient in GAMESS US computer program\cite{GAMESS.JCC.1993}
with canonical molecular orbitals
is on average 20--30 times more demanding than the 
other components,
%exchange\hyp{}repulsion
%component, 
becoming a bottleneck of the evaluation of total interaction 
energy and gradients.\cite{Gordon.Smith.Xu.Slipchenko.AnnuRevPhysChem.2013,
Li.Gordon.Jensen.JCP.2006}
Recent advancement 
of Xu and Gordon\cite{Xu.Gordon.JCP.2013} 
reduced the cost further by about 50\% 
by minimizing the size of the virtual orbital space
via the use of quasiatomic minimal\hyp{}basis orbitals\cite{Lu.Wang.Schmidt.Bytautas.Ho.Reudenberg.JCP.2004} (QUAMBO's).
Unfortunately, even with this improvement,
%this was still insufficient to overcome
%the high cost of evaluation of the potential energy integrals, making 
the CT term remains still the most time\hyp{}consuming
to evaluate from among all the EFP2 terms.

In effect, the CT energy component is often ignored 
in some applications.\cite{Kuroki.Mori.JPCB.2019,
Blasiak.Londergan.Webb.Cho.ACR.2017,
Tommaso.Lafiosca.Cappelli.JCTC.2017,
Gurunathan.Acharya.Ghosh.Kosenkov.Kaliman.Shao.Krylov.Slipchenko.JPCB.2016,
Budzak.Laurent.Laurence.Medved.Jacquemin.JCTC.2016,
Smith.Gordon.Slipchenko.JPCA.2011,
Smith.Gordon.Slipchenko.JPCA.2011.2,
Ghosh.Kosenkov.Vanovschi.Williams.Herbert.Gordon.Schmidt.Slipchenko.Krylov.JPCA.2010,
Smith.Gordon.Slipchenko.JPCA.2008}
In fact,
EFP2 CT term is available only in the GAMESS US quantum chemistry program,\cite{GAMESS.JCC.1993}
whereas it is neither supported in the official release 
of the recent LIBEFP library for linking quantum chemistry packages 
with the EFP2 functionalities,\cite{Kaliman.Slipchenko.JCC.2015}
nor in the Q\hyp{}CHEM quantum chemistry 
program,\cite{Ghosh.Kosenkov.Vanovschi.Flick.Kaliman.Shao.Gilbert.Krylov.Slipchenko.JCC.2013}
contrary to electrostatic, exchange\hyp{}repulsion, induction and dispersion EFP2 terms.

The effect of CT on the interaction energy 
is known to be often non\hyp{}negligible, especially
in donor\hyp{}acceptor systems such as H\hyp{}bonded species
and charged complexes,\cite{Devarajan.Gaenko.Gordon.Windus.Fragmentation.2017,
Devarajan.Windus.Gordon.JPCA.2011} 
although it highly depends on the theoretical
approach being used, i.e., the choice of the reference
wavefunctions.
The approach adopted by the founders of EFP2 model
is based on gas\hyp{}phase reference (unperturbed) HF wavefunctions,
which will be here broadly referred to as the CT/HF0 level of theory
and the symbol `0' denotes unperturbed wavefunctions.
In this formulation, the CT energy can be understood
as a part of the polarization energy that cannot be
linked to dispersion and pure induction interactions. 
In technical words, mixing of virtual and occupied MO's,
that are assigned according to certain criteria 
to different unperturbed wavefunctions, occurs
unlike the pure induction and dispersion, in which such mixings are contained
solely within one monomer.\cite{Stone.TheTheoryOfIntermolecularForces.1996}

One of the main goals of this work, apart from developing
more efficient model of the CT/HF0 energy that
is compatible with the EFP2 energetics, is to
extend the one\hyp{}electron effective potentials (here referred to as the OEP) 
technique
widely
used previously\cite{Roothaan.RevModPhys.1951,
Hohenberg.Kohn.PhysRev.1964,
Kohn.Sham.PhysRev.1965,
Otto.Ladik.ChemPhys.1975,
Holas.March.PhysRevA.1991,
Weber.Thiel.TCA.2000,
Neese.JCP.2005,
Cisneros.Andres.Piquemal.Darden.JCP.2005,
Piquemal.Cisneros.Reinhardt.Gresh.Darden.JCP.2006,
Li.Gordon.Jensen.JCP.2006,
Blasiak.Lee.Cho.JCP.2013,
Blasiak.Maj.Cho.Gora.JCTC.2015}
to simplify the rigorous and costly quantum chemistry models of extended systems
with a particular emphasis on solvation phenomena and molecular dynamics.
The presented OEP technique of removing ERI's from the working
equations
follows the notion of the importance of 
one\hyp{}electron densities in chemistry\cite{Kohn.Sham.PhysRev.1965,Holas.March.PhysRevA.1991},
thus reducing the complicated summations involving ERI's to much shorter expressions involving
only one\hyp{}electron integrals (OEI's).
Therefore, the OEP computational method is first presented in Section~\ref{s:2.oep}.
Next, in Section~\ref{s:3.ct},
the new technique is used to derive an alternative formulation of the CT/HF0 energy
compatible with the EFP2 method. Subsequently, after details of computations
are discussed in Section~\ref{s:4.calculations}, in Section~\ref{s:5.results} the validation of the 
OEP\hyp{}based CT/HF0 model is presented and its performance in terms 
of accuracy and computational speed is compared against the EFP2
model. Finally, few concluding remarks and outlook of future work
are given in Section~\ref{s:6.conclusions}.

\section{\label{s:2.oep}Effective One-Electron Potential Operators}

To establish the notation within the present work, 
the formalism of one\hyp{}electron potential operators is
given first.
The one\hyp{}electron Coulomb static effective potential $v^{\rm eff}({\bf r})$
produced by a certain effective one\hyp{}electron charge density distribution $\rho^{\rm eff}({\bf r})$
is 
\begin{equation} \label{e:v-eff}
	v^{\rm eff}({\bf r}) = \int \frac{ \rho^{\rm eff}({\bf r}') }{ \vert {\bf r}' - {\bf r} \vert} d{\bf r}' \;,
\end{equation}
where ${\bf r}$ is a spatial coordinate. 
For convenience, the total effective density can be split into nuclear and electronic contributions,
\begin{equation} \label{e:rho-eff}
 \rho^{\rm eff}({\bf r}) = \lambda \rho^{\rm eff}_{\rm nuc}({\bf r}) + \rho^{\rm eff}_{\rm el}({\bf r}) \;,
\end{equation}
where $\lambda$ is a certain parameter and is assumed to be either 1 or 0 in this work.
Formally, the electronic part can be expanded in terms of an effective
bond order matrix, or one\hyp{}particle density (OED), represented in a certain basis of orbitals, $\BM{\phi}({\bf r})$,
\begin{equation} \label{e:d-spectral}
	\rho^{\rm eff}_{\rm el}({\bf r}) = -\sum_{\alpha\beta} P_{\alpha\beta}^{\rm eff} 
	\phi_\alpha({\bf r}) \phi_\beta^{*}({\bf r})  \;.
\end{equation}
Based on that, the operator form of the effective potential 
can be written as
\begin{equation} \label{e:oep-operator}
	\hat{v}^{\rm eff} = 
        \lambda \hat{v}_{\rm nuc} +
        \int d{\bf r} \Ket{{\bf r}} 
        v^{\rm eff}_{\rm el}({\bf r})
        \Bra{{\bf r}}
\end{equation}
with the nuclear
operator
defined by
\begin{equation} \label{e:oep-operator-nuc}
        \hat{v}_{\rm nuc} \equiv \sum_x^{\rm At}  
                     \int d{\bf r} \Ket{{\bf r}} 
                     \frac{Z_x}{\vert {\bf r} - {\bf r}_x \vert}
                     \Bra{{\bf r}} \;.
\end{equation}
and $\BraKet{\bf r}{\alpha} \equiv \varphi_\alpha({\bf r})$.
The matrix element of the effective potential operator
is therefore given by
\begin{equation} \label{e:oep-matrix-element}
	\tBraKet{\alpha}{ \hat{v}^{\rm eff} }{\beta}
	= \lambda \sum_x^{\rm At} W_{\alpha\beta}^{(x)} -
        \sum_{\gamma\delta} \BraKet{\alpha\beta}{\gamma\delta} P^{\rm eff}_{\gamma\delta}  \;,
\end{equation}
where 
\begin{equation} \label{e:oep-matrix-element-nuc}
 W_{\alpha\beta}^{(x)} = 
 Z_x \int \frac{\phi^*_\alpha({\bf r}) \phi_\beta({\bf r})}{\vert {\bf r} - {\bf r}_x \vert} d{\bf r} \;,
\end{equation}
$Z_x$ is the atomic number of the $x$th atom,
and the symbol `At' denotes all atoms that contribute to the effective potential.
In the above equation and throughout the work, 
\begin{equation} \label{e:1el-operator}
\tBraKet{\alpha}{\mathscr{O}(1)}{\beta} \equiv \int \phi^*_\alpha({\bf r}) \mathscr{O}(1) \phi_\beta({\bf r}) d{\bf r} 
\end{equation}
for any one\hyp{}electron operator $\mathscr{O}(1)$ 
and the ERI
is defined according to
\begin{equation} \label{e:eri}
	\BraKet{\alpha\beta}{\gamma\delta} \equiv
	\iint 
	\frac{ \phi_\alpha^{*}({\bf r}_1) \phi_\beta({\bf r}_1) 
	       \phi_\gamma^{*}({\bf r}_2) \phi_\delta({\bf r}_2) }{ \vert {\bf r}_1 - {\bf r}_2 \vert}
	d{\bf r}_1 d{\bf r}_2  \;.
\end{equation}

\subsection{\label{ss:2.1.oep-technique}Incorporating Electron Repulsion Integrals into Effective Potentials}

Consider now an arbirtary functional $\mathcal{F}$ that explicitly depends on the 
ERI's. In this work, OEP's are defined by the following transformation
 \begin{equation}
 \mathcal{F}
 \left[ 
   \BraKet{ij}{k^Al^A}
	 \right] = \tBraKet{i}{\hat{v}_{kl}^A}{j}  \;,
 \end{equation}
where 
$\hat{v}_{kl}^A$ is the effective OEP operator given by Eq.~\eqref{e:oep-operator} 
with the
effective density $\rho_{kl}^A({\bf r}) \equiv \phi_k^A({\bf r})\phi_l^A({\bf r})$.
The summations over $k$ and $l$ can be incorporated into the total effective OEP operator
$\hat{v}_{\text{eff}}^A$
to produce
\begin{equation}
	\sum_{ij}\sum_{kl\in A} \mathcal{F}\left[ 
   \BraKet{ij}{k^Al^A}
 \right] = \sum_{ij} \tBraKet{i}{\hat{v}_{\text{eff}}^A}{j}  \;.
\end{equation}
Thus, the total computational effort is, in principle, reduced from the fourth\hyp{}fold
sum involving evaluation of ERI's to the two\hyp{}fold sums of cheaper OEI's.
It is also possible to generalize the above expression even further by
summing over all possible functionals ${\mathcal{F}}_t$
\begin{equation} \label{e:ft-reduction}
	\sum_t \sum_{ij}\sum_{kl\in A} {\mathcal{F}}_t\left[ 
   \BraKet{ij}{k^Al^A}
 \right] = \sum_{ij} \tBraKet{i}{\hat{v}_{\text{eff}}^A}{j} \;.
\end{equation}
The above design has the advantage that it opens the possibility to define first\hyp{}principles
effective fragments as long as 
the functionals ${\mathcal{F}}_t$ 
are well defined, computable and can be approximately
partitioned in between the interacting fragments.

Three unique classes of ERI's can be recognized 
based on the basis function partitioning scheme
within the system composed of two molecules (shall be $A$ and $B$
throughout the course of this work). 
They are as follows:
\begin{enumerate}
\item the Coulomb\hyp{}like ERI's of the type 
$\BraKet{AA}{BB}\rightarrow \BraKet{\phi_{i\in A}\phi_{k\in A}}{\phi_{j\in B}\phi_{l\in B}}$,
\item the overlap\hyp{}like ERI's of the type 
$\BraKet{AA}{AB}\rightarrow \BraKet{\phi_{i\in A}\phi_{k\in A}}{\phi_{j\in A}\phi_{l\in B}}$, and
\item the exchange\hyp{}like ERI's of the type 
$\BraKet{AB}{AB}\rightarrow \BraKet{\phi_{i\in A}\phi_{j\in B}}{\phi_{k\in A}\phi_{l\in B}}$.
\end{enumerate}
In contrast to 
the first two classes of ERI's, 
exchange\hyp{}like ERI's cannot be incorporated into OEP's. 
The Coulomb and overlap\hyp{}like classes, which are listed in Table~\ref{t:oep-matrix-element-types},
are usually approximated via
expanding the OEP operator in distributed multipole (DMTP) expansion series, and integrating over
one remaining electron coordinate,\cite{Li.Gordon.Jensen.JCP.2006} i.e.,
\begin{multline} \label{e:coulomb-energy:dmtp-oei}
 \tBraKet{j}{\hat{v}_{\rm eff}^A}{l} 
 \cong  
 \Bigg< j \Bigg|  \sum_{w\in A} \Big\{
                                 T^{(0)}_{we} \hat{q}_w 
                          - {\bf T}^{(1)}_{we} \cdot \hat{\BM\upmu}_w  \\
            + \frac{1}{3}   {\bf T}^{(2)}_{we} :     \hat{\BM\Theta}_w 
                    - \ldots  
                \Big\}
              \Bigg| l \Bigg>_{{\bf r}_e} \;.
\end{multline}
In the above equation, $\hat{q}_{\rm eff}^{(w)}$,
$\hat{\BM\upmu}^{(w)}_{\rm eff}$ and $\hat{\BM\Theta}^{(w)}_{\rm eff}$
are the quantum operators of effective distributed monopole (charge),
dipole moment and quadrupole moment, respectively, centered on the $w$th site at ${\bf r}_w$, 
whereas ${{\bf T}^{(d)}_{we}}$ are the so called interaction tensors of rank $d$ between ${\bf r}_w$ and ${\bf r}_e$, with the latter being the electronic coordinate.
Explicit forms of interaction tensors can be found elsewhere.\cite{Blasiak.Cho.JCP.2014}
Note that $\Ket{j}$ and $\Ket{l}$ can belong to either molecule.
In this way, ERI's are no longer needed and the computational cost reduces appreciably.
In this work, this method is however considered already too expensive
for application in the CT/HF0 energy because evaluation of
Eq.~\eqref{e:coulomb-energy:dmtp-oei} requires calculation
of electrostatic potential and electrostatic potential gradient(s) OEI's. These kind of
integrals are typically the most expensive when compared to other standard OEI's
such as overlap or kinetic energy integrals.
Therefore, Coulomb and overlap\hyp{}like OEP matrix elements
will be treated via more approximate and less expensive approaches
which are discussed next.

{
\renewcommand{\arraystretch}{1.4}
\begin{table}[b]
\caption[Types of matrix elements with OEP operators]
{{\bf Types of matrix elements with OEP operators}
}
\label{t:oep-matrix-element-types}
\begin{ruledtabular}
\begin{tabular}{lcccc}
Matrix element      &&            Overlap-like                  &&             Coulomb-like                \\ 
                    && $\tBraKet{i}{\hat{v}^{A}_{\rm eff}}{j} $ && $\tBraKet{j}{\hat{v}^{A}_{\rm eff}}{l}$ \\ 
	\cline{1-5}
Partitioning scheme &&            $i\in A, j\in B$              &&               $j,l\in B$                \\
ERI class           &&            $\BraKet{AA}{AB}$             &&               $\BraKet{AA}{BB}$         \\
DF\footnotemark[1]/RI\footnotemark[2] Form    
&& $\sum_{\xi\in A}^{\rm } v^A_{i\xi} S^{AB}_{\xi j} $  
&& $\sum_{\xi\zeta\in A}^{\rm } S^{BA}_{j\xi} v^A_{\xi\zeta} S^{AB}_{\zeta l} $ \\
DMTP\footnotemark[3] Form                     
&& --  &  &  $\rho_{jl}^B \odot \rho_{\rm eff}^A$ \\
\end{tabular}
\end{ruledtabular}
\footnotetext[1]{Density Fitting}
\footnotetext[2]{Resolution of Identity}
\footnotetext[3]{Distributed Multipole Expansion}
\end{table}
}

\subsection{\label{ss:2.2.oep-DMTP}Semi-classical multipole expansion}

Semi\hyp{}classical multipole expansion is the most applicable in case of matrix elements of the type
$
 \tBraKet{j^B}{\hat{v}_{\rm eff}^A}{l^B}
$
because it can be considered as a Coulombic interaction between $\rho^B_{jl}$
and $\rho^A_{\rm eff}$. These matrix elements require ERI's of type
$\BraKet{AA}{BB}$ only.
In general, given certain two effective one\hyp{}electron density distributions,
the associated effective Coulombic interaction energy can be estimated from the classical formula
according to
\begin{equation} \label{e:coulomb-energy:general}
  E_{\rm eff} = \iint \frac{\rho_{\rm eff}^X({\bf r}_1) \rho_{\rm eff}^Y({\bf r}_2)}
 {\vert {\bf r}_1 - {\bf r}_2 \vert} 
d{\bf r}_1 d{\bf r}_2 \;.
\end{equation}
The above integral can be approximated by 
applying the DMTP expansion 
to both effective potential operators
and dropping the $\hat{}$ symbol in the multipole operators
which leads to:\cite{Stone.TheTheoryOfIntermolecularForces.1996}
\begin{multline} \label{e:coulomb-energy:dmtp}
  E_{\rm eff} \approx
 \rho_{\rm eff}^X \odot \rho_{\rm eff}^Y \equiv
 \sum_{u\in A} \sum_{w\in B} \Big\{ 
 T^{(0)}_{uw}
 q_{\rm eff}^{(u)}  q_{\rm eff}^{(w)} \\
 - {\bf T}^{(1)}_{uw} \cdot 
   \left[ q_{\rm eff}^{(u)} {\BM\upmu}_{\rm eff}^{(w)} - q_{\rm eff}^{(w)} {\BM\upmu}_{\rm eff}^{(u)} \right] 
 - {\bf T}^{(2)}_{uw} : 
  {\BM\upmu}_{\rm eff}^{(u)}  \otimes {\BM\upmu}_{\rm eff}^{(w)} 
 \ldots
 \Big\}
\end{multline}
The symbol `$\odot$' denotes the sum of all the tensor contractions
performed over the DMTP's of molecule $X$ and $Y$ to yield the associated interaction energy.
The choice of the distribution centres as well as the truncation order of the multipole expansion
is crucial in compromising the accuracy and computational cost of the resulting expressions.
There are many ways in which this can be achieved, e.g., through the distributed multipole analysis (DMA)
of Stone and Alderton\cite{Stone.Alderton.MolPhys.1985,Stone.JCTC.2005},
the cumulative atomic multipole moments (CAMM) of Sokalski and Poirier\cite{Sokalski.Poirier.CPL.1983},
the localised distributed multipole expansion (LMTP) of Etchtebest, Lavery and Pullman\cite{Etchebest.Lavery.Pullman.TheorChimActa.1982}
or other schemes based on fitting to electrostatic potential, such as ChelpG\cite{Breneman.Wiberg.JCC.1990}.

\subsection{\label{ss:2.3.oep-GDF}Extended Density Fitting of OEP's}

Extended density fitting of OEP's, 
which will be referred to as the EDF scheme, 
is applicable in case of matrix elements of 
$
 \tBraKet{i^A}{\hat{v}_{\rm eff}^A}{j^B}
$ type. 
 These matrix elements require ERI's of type
$\BraKet{AA}{AB}$ only.
To get the \emph{ab initio} representation of such an overlap\hyp{}like matrix element,
one can use a procedure similar to
the typical density fitting (DF) or resolution of identity (RI), which are nowadays widely used 
to compute electron\hyp{}repulsion integrals (ERI's) more efficiently,
and reduce computational cost of post\hyp{}Hartree\hyp{}Fock methods.\cite{Hesselmann.Jansen.Schutz.JCP.2005} Density fitting was also
applied to design \emph{ab initio} force fields.\cite{Piquemal.Cisneros.Reinhardt.Gresh.Darden.JCP.2006,Cisneros.Andres.Piquemal.Darden.JCP.2005}

\subsubsection{\label{sss:2.3.1.GDF-1}Density Fitting in Nearly-Complete Space}

An arbitrary one\hyp{}electron potential of molecule $A$ acting on any state vector 
associated with molecule $A$ can be expanded in an auxiliary space centered 
on $A$ as
\begin{equation}
   \hat{v}^{A}\Ket{i} = \sum_{\xi\eta}^{\rm RI} \hat{v}^{A}\Ket{\xi} [{\bf S}^{-1}]_{\xi\eta} \BraKet{\eta}{i}
\end{equation}
under the necessary assumption that the auxiliary basis set is nearly complete,
i.e., 
$\sum_{\xi\eta}^{\rm RI} \Ket{\xi}[{\bf S}^{-1}]_{\xi\eta} \Bra{\eta} \cong 1$.
In the equations above, 
the `RI' symbol denotes a certain auxiliary basis set that fulfills such resolution of identity. 
The above general expansion can be also obtained by 
utilizing the density fitting
in the nearly\hyp{}complete space,
\begin{equation} \label{e:v-compl}
 \hat{v}^{A}\Ket{i} = \sum_{\xi}^{\rm RI} V^A_{i\xi} \Ket{\xi} \;.
\end{equation}
In the above equation,
the matrix ${\bf V}^A$
is the projection of the state vector $\hat{v}^{A} \Ket{i}$
onto the nearly\hyp{}complete basis $\{ \varphi_\xi \}$.
Let $Z_i[{\bf V}^A]$ be the least\hyp{}squares objective function 
\begin{equation} \label{e:z-compl}
 Z_i[{\bf V}^A] = \int \vert \Xi_i({\bf r}) \vert^2 d{\bf r}
\end{equation}
with the error density defined by
\begin{equation} \label{e:error-compl}
 \Xi_i({\bf r}) = v^A({\bf r}) \phi_i({\bf r}) - \sum_\xi^{\rm RI} V^A_{i\xi} \varphi_\xi({\bf r}) \;.
\end{equation}
By requiring that
\begin{equation} \label{e:z-necessary-requirement}
 \frac{\partial Z_i[{\bf V}^A]}{\partial V^A_{i\mu}} = 0 \text{ for all $\mu$}
\end{equation}
one finds the coefficients of the $i$th row of ${\bf V}^A$ to be
\begin{equation} \label{e:gdf-compl.v}
% {\bf v}_i = {\bf a}^{(i)} \cdot {\bf S}^{-1}
  V^A_{i\xi} = \sum_\eta^{\rm RI} [{\bf S}^{-1}]_{\xi\eta} a^{(i)}_\eta \;,
\end{equation}
where the auxiliary matrices are given by
\begin{subequations}
\begin{align}
 a^{(i)}_\eta &= \int \varphi^*_\eta({\bf r}) \hat{v}^A \phi_i({\bf r}) d{\bf r} \;,\\  
 S_{\eta\xi}  &= \int \varphi^*_\eta({\bf r}) \varphi_\xi({\bf r}) d{\bf r} \;.
\end{align}
\end{subequations}
The working formula for $a^{(i)}_\eta$ can be found by applying 
potential form from Eq.~\eqref{e:v-eff}
along with the spectral representation of the effective density from Eq.~\eqref{e:d-spectral} 
which finally gives
\begin{equation} \label{e:a-coeff-compl}
 a^{(i)}_\eta = \lambda \sum_{x} W_{\eta i}^{(x)} - 
 \sum_{\alpha\beta} P^A_{\alpha\beta} 
  \BraKet{\alpha\beta}{\eta i} \;.
\end{equation}
Since only one step is required to obtain the OEP matrix, the scheme will
be abbreviated as EDF-1.

\subsubsection{\label{sss:2.3.1.GDF-2}Density Fitting in Incomplete Space}

Density fitting scheme from previous section has practical disadvantage of a nearly\hyp{}complete basis set
being usually very large (spanned by large amount of basis set vectors). 
Since most of basis sets used in quantum chemistry do not form a nearly complete
set, it is beneficial to design a modified scheme in which it is possible to obtain the effective 
matrix elements of the OEP operator in an incomplete auxiliary space. This can be achieved by minimizing 
the following objective function\cite{Cisneros.Andres.Piquemal.Darden.JCP.2005,Piquemal.Cisneros.Reinhardt.Gresh.Darden.JCP.2006}
\begin{equation} \label{e:z-incompl}
	Z_i[{\bf V}^A] = \iint 
        \frac{ \Xi_i^*({\bf r}_1) \Xi_i({\bf r}_2) }{\vert {\bf r}_1 - {\bf r}_2 \vert}  
         d{\bf r}_1 d{\bf r}_2  \;.
\end{equation}
with the error density defined by
\begin{equation} \label{e:error-incompl}
 \Xi_i({\bf r}) = v^A({\bf r}) \phi_i({\bf r}) - \sum_\xi^{\rm DF} V^A_{i\xi} \varphi_\xi({\bf r}) \;.
\end{equation}
where the symbol `DF' denotes the generally incomplete auxiliary basis set.
From the requirement given in Eq.~\eqref{e:z-necessary-requirement}
one obtains
\begin{equation} \label{e:v-incompl}
% {\bf G}^{(i)} = {\bf b}^{(i)} \cdot {\bf A}^{-1}
  V^A_{i\xi} = \sum_\eta^{\rm DF} [{\bf R}^{-1}]_{\xi\eta} b^{(i)}_\eta \;,
\end{equation}
where 
\begin{subequations}
\begin{align}
 b^{(i)}_\eta &= \iint 
                       \frac{ \varphi^*_\eta({\bf r}_1) \hat{v}^A \phi_i({\bf r}_2) } 
                            {\vert {\bf r}_1 - {\bf r}_2\vert}  
                 d{\bf r}_1 d{\bf r}_2 \;, \\
 R_{\eta\xi}  &= \iint 
                       \frac{ \varphi^*_\eta({\bf r}_1) \varphi_\xi({\bf r}_2) } 
                            {\vert {\bf r}_1 - {\bf r}_2\vert}  
                 d{\bf r}_1 d{\bf r}_2 \;.
\end{align}
\end{subequations}
Note that, while $R_{\eta\xi}$ is a typical 2\hyp{}center ERI 
that can be evaluated by standard means,\cite{McMurchie.Davidson.JComputPhys.1978}
$b^{(i)}_\eta$ matrix elements are not trivial to calculate
because the OEP operator, which contains integration over an electron coordinate,
is present inside the double integral. Therefore, the following triple integral,
\begin{equation} \label{e:triple-integral}
 b^{(i)}_\eta = \iiint 
           \frac{ \varphi^*_\eta({\bf r}_1) \phi_i({\bf r}_2)  \rho^{\rm eff}({\bf r}_3) }
            {\vert {\bf r}_1 - {\bf r}_2 \vert \vert {\bf r}_3 - {\bf r}_2 \vert}
           d{\bf r}_1 d{\bf r}_2 d{\bf r}_3 \;,
\end{equation}
has to be computed.
Obtaining all the necessery integrals of this kind directly 
by performing integrations of Eq.~\eqref{e:triple-integral} is very costly 
even for medium sized molecules.\cite{Barca.Loos.JCP.2017}
However, one can introduce the effective potential in order to eliminate one integration. 
This can be achieved by performing additional density fitting 
in a nearly complete intermediate basis\cite{Barca.Loos.JCP.2017} 
\begin{equation}
 \hat{v}^{\rm eff} \Ket{i} = \sum_\varepsilon^{\rm RI} H_{i\varepsilon} \Ket{\varepsilon} \;,
\end{equation}
The working equation is therefore given by
\begin{equation} \label{e:v-compl-b-coeff}
 b^{(i)}_\eta = \sum_\varepsilon^{\rm RI} H_{i\varepsilon} R_{\varepsilon\eta} \;,
\end{equation}
which can be easily evaluated by noting that
$H_{i\varepsilon}$ components 
are given by Eq.~\eqref{e:v-compl}
--- thus, the matrix elements of the OEP operator can be found
by using Eq.~\eqref{e:v-incompl}.
It is emphasized here that, as long as the state vector $\Ket{i}$, OEP operator $\hat{v}^A$, and
the auxiliary and intermediate basis sets depend solely on the unperturbed molecule $A$, the matrix elements
$V^A_{i\xi}$ can be calculated just once and stored as effective fragment parameters.

In contrast to the EDF-1 scheme, two density fitting steps are now required to
obtain the OEP matrix. Thus, extended density fitting in incomplete basis will be referred to as the EDF-2 scheme.
In the limitting case, when the auxiliary basis is the same as the intermediate basis,
EDF-2 is equivalent to EDF-1.
As the theory required to define OEP's and their computational
evaluation in the fragment\hyp{}based formulation has been finally given in this Section,
the application of the OEP technique of ERI elimination is
further discussed for the evaluation of the CT/HF0 energy in the following sections.

\section{\label{s:3.ct}Charge Transfer Interaction Energy for Fragment Potentials}

In the CT/HF0 treatments of bi\hyp{}molecular complexes, the CT energy
can be expressed as a sum of the energy stabilization due to 
CT from molecule $A$ to $B$ and vice versa, i.e.,
\begin{equation} \label{e:ct-hf0}
 E^{\rm CT} =  E^{A\rightarrow B} + E^{B\rightarrow A}\;.
\end{equation}

\subsection{\label{ss:3.1.EFP2} EFP2 Model}

Li, Gordon and Jensen used the expansion of the overlap density 
in Taylor series and found four different approximate
theories for the CT energy.\cite{Li.Gordon.Jensen.JCP.2006} 
The optimal theory, which was shown
to well reproduce the CT/HF0 energies obtained by using 
the reduced variational space (RVS)
analysis of Stevens and Fink\cite{Stevens.Fink.CPL.1987},
reads as
\begin{equation} \label{e:u-efp2-ct}
 E^{A\rightarrow B} \approx 
 2\sum_{i\in A}^{\rm Occ}
  \sum_{n\in B}^{\rm Vir}
   \frac{ \left| U_{in}^{A\rightarrow B} \right|^2}{\varepsilon_{i} - T_{nn}} \;,
\end{equation}
where
\begin{multline} \label{e:u-efp2-ct-v2}
 \left| U_{in}^{A\rightarrow B} \right|^2 \approx
 \frac{u_{in}}{1-\sum_{m\in A}^{\rm All} S_{mn}^2}
 \Bigg\{
   u_{in} \\+ \sum_{j\in B}^{\rm Occ} S_{ij} 
   \left( 
     T_{nj} - \sum_{m\in A}^{\rm All} S_{nm} T_{mj}
   \right)
 \Bigg\} \;,
\end{multline}
and 
\begin{equation} \label{e:uin-efp2}
 u_{in} \equiv U_{in}^{{\rm EF},B} - \sum_{m\in A}^{\rm All} U_{im}^{{\rm EF},B} S_{mn} \;,
\end{equation}
in which the summations extend over occupied (denoted by `Occ'), virtual (denoted by `Vir') 
or both (denoted by `All') canonical HF MO's. The effective potential energy
matrix elements are defined by
\begin{equation} \label{e:v-efp2-eff}
 U_{in}^{{\rm EF},B} \equiv 
 -\tBraKet{i}{\hat{v}^B_{\rm tot} }{n} \;,
\end{equation}
and are evaluated by expanding the $\hat{v}^B_{\rm tot}$ operator
in distributed multipoles according to Eq.~\eqref{e:coulomb-energy:dmtp-oei}.
The apparent success of the EFP2 scheme is rooted in the dramatic simplifications of the
\emph{ab initio} expressions for interaction energy, in which
the relatively costly ERI's
have been effectively removed from the working models while maintaining the required accuracy. 
That is, in the case of the CT/EFP2 energy component, 
the canonical orbital energies $\varepsilon_i$ are
constant parameters, whereas the overlap $S_{nm}$, kinetic energy $T_{mn}$ and effective
one\hyp{}electron electrostatic potential $U_{in}^{{\rm EF}}$ matrix elements
are all certain types of the one\hyp{}electron integrals, orders of magnitude cheaper to
evaluate than ERI's. Unfortunately, due to extensive summations over virtual orbitals,
evaluating Eq.~\eqref{e:u-efp2-ct} is still considerable in cost because typically large basis sets
need to be used for generating the EFP2 parameters. In effect, calculation of $U_{in}^{{\rm EF}}$
is much more expensive as compared to other types of one\hyp{}electron integrals
and is the bottleneck of CT/EFP2 energy calculation,
even when using minimal amount of virtual orbitals.\cite{Xu.Gordon.JCP.2013} 

In the following subsections, the alternative model of the CT/HF0 energy is proposed
by introducing OEP's. Although application of the OEP method to the CT/HF0 formulation in EFP2 method
is probably possible, it would be relatively difficult to discuss the resulting OEP\hyp{}based
EFP2 models because there are four distinct versions of this theory
with a set of different approximations, selected based on performance assessment rather than 
a rigorous derivation manner.\cite{Li.Gordon.Jensen.JCP.2006}
Instead, perturbation theory of Murrell et al.\cite{Murrell.Randic.Williams.Longuet-Higgins.ProcRSocLondA.1965} 
with the
explicit formulation for closed shell systems by Otto and Ladik\cite{Otto.Ladik.ChemPhys.1975},
which is somewhat more rigorous than the EFP2 CT model,
is chosen as a starting point in this work. 
It is believed that this choice will enable a clear demonstration 
of the OEP technique in fragment\hyp{}based modelling.

\subsection{\label{ss:3.2.OL} Otto-Ladik's Model: Starting Point}

The CT/HF0 energy can be expressed
by\cite{Murrell.Randic.Williams.Longuet-Higgins.ProcRSocLondA.1965,Otto.Ladik.ChemPhys.1975}
\begin{equation} \label{e:ct-murell-etal.e}
 E^{A\rightarrow B} = 2 \sum_{i\in A}^{\rm Occ} \sum_{n\in B}^{\rm Vir} 
  \frac{\vert U^{A\rightarrow B}_{in} \vert^2 }{\varepsilon_i - \varepsilon_n} \;,
\end{equation}
where the coupling constant is given by Otto and Ladik,\cite{Otto.Ladik.ChemPhys.1975}
here referred to as the OL method, as
\begin{multline} \label{e:ct-murell-etal.vin}
 U^{A\rightarrow B}_{in} = 
      - \tBraKet{i}{\hat{v}^B_{\rm tot} }{n} 
      - \sum_{j\in B}^{\rm Occ} \BraKet{nj}{ij} \\
      + \sum_{k\in A}^{\rm Occ} S_{nk} \tBraKet{k}{\hat{v}^B_{\rm tot} }{i} 
      + \sum_{j\in B}^{\rm Occ} S_{ij} \tBraKet{j}{\hat{v}^A_{i} }{n}  \\
     + \sum_{k\in A}^{\rm Occ} \sum_{j\in B}^{\rm Occ}  
        S_{kj} \left(1+\delta_{ik}\right) \BraKet{nj}{ik} \;.
\end{multline}
In the above expression, the following effective one\hyp{}electron potential operators,
\begin{subequations} 
\begin{align} \label{e:ct-murell-etal.vtot-vi}
 \hat{v}^B_{\rm tot} &= \hat{v}^B_{\rm nuc} + 2\sum_{j\in B}^{\rm Occ} \hat{v}^B_{jj} \quad\text{ and }\\ 
 \hat{v}^A_{i      } &= \hat{v}^A_{\rm tot} - 2\hat{v}^A_{ii} \;,
\end{align}
\end{subequations}
were introduced without making any approximation to the original equation
from Ref.\cite{Otto.Ladik.ChemPhys.1975} 
Note that ERI's in MO basis are necessary to evaluate all terms
in Eq.~\eqref{e:ct-murell-etal.vin}, e.g., 
$\tBraKet{i}{\hat{v}^B_{\rm tot} }{n} \equiv W^B_{nj} - 
2\sum_{j\in B}\BraKet{in}{jj}$.
Therefore, the goal of this work is to apply the technique sketched in Eq.~\eqref{e:ft-reduction}
to effectively eliminate ERI's from Eq.~\eqref{e:ct-murell-etal.vin} in terms of OEP's.

\subsection{\label{ss:3.3.OEP} Otto-Ladik's Model: Application of OEP Technique}

One can immediately notice that the five summation terms
from Eq.~\eqref{e:ct-murell-etal.vin} can be classified 
based on Table~\ref{t:oep-matrix-element-types} 
into three groups
regarding the type of ERI's that are required:
(i) overlap\hyp{}like $\BraKet{AB}{BB}$ -- the first two terms;
(ii) Coulomb\hyp{}like $\BraKet{AA}{BB}$ -- the third term and
(iii)  Coulomb\hyp{}like $\BraKet{BB}{AA}$ -- the two last terms. 
Note also that there are no exchange\hyp{}like terms needed in this case.
Therefore, all the contributions can be re\hyp{}cast in terms of the OEP's.

\paragraph{Group (i).}
Group (i) can be rewritten by noticing that
\begin{equation} \label{e:ct-murell-etal.group-i.notice}
 \sum_{j\in B}^{\rm Occ} 
 \BraKet{nj}{ij} =
 \Bra{i} \left[ -\hat{v}^B_{nj} \Ket{j} \right]  \;,
\end{equation}
which, by combining with the first term from Eq.~\eqref{e:ct-murell-etal.vin}, 
allows to apply the two\hyp{}step extended density fitting (EDF-2) 
developed in Section~\ref{sss:2.3.1.GDF-2} as follows
\begin{equation} \label{e:v-oep.ct}
\Bra{i} \left[ -\hat{v}^B_{\rm tot} \Ket{n} + \sum_{j\in B}^{\rm Occ} \hat{v}^B_{nj} \Ket{j} \right]
\cong \Bra{i} \sum_{\eta\in B}^{\rm DF} V^B_{n\eta} \Ket{\eta}
\end{equation}
with the error density given by
\begin{multline}
 \Xi_n({\bf r}) = 
  \left\{
      v_{nuc}^B({\bf r}) + 2 \sum_{j\in B}^{\rm Occ}
   \int \frac{ \vert \phi_j({\bf r}') \vert^2 }{\vert {\bf r}' - {\bf r} \vert} d{\bf r}'
  \right\}
  \phi_n({{\bf r}})  \\
 - \sum_{j\in B}^{\rm Occ} \phi_j({{\bf r}})
   \int \frac{ \phi^*_n({\bf r}') \phi_j({{\bf r}'}) }{\vert {\bf r}' - {\bf r} \vert} d{\bf r}' 
 - \sum_{\eta\in B}^{\rm DF} V^B_{n\eta} \varphi_\eta({\bf r}) \;.
\end{multline}
Substituting the above equation into Eq.~\eqref{e:z-incompl}
leads to
\begin{equation} \label{e:v-oep-group-1-prefinal}
 V^B_{n\xi} = \sum_{\eta\in B}^{\rm DF} \left[ {\bf R}^{-1} \right]_{\xi\eta} b_\eta^{(n)} \;,
\end{equation}
where
\begin{equation}
 b^{(n)}_\eta = \iint 
           d{\bf r}_1 d{\bf r}_2  
           \frac{ \varphi^*_\eta({\bf r}_1) }
            {\vert {\bf r}_1 - {\bf r}_2 \vert }  \times
          \left[ 
           \hat{v}^B_{\rm tot}  \phi_n({\bf r}_2) 
         - \sum_{j\in B}^{\rm Occ} \hat{v}^B_{nj} \phi_j({{\bf r}_2})
           \right]  \;.
\end{equation}
The above result is given by a sum of triple integrals from Eq.~\eqref{e:triple-integral}.
However, the following application of the resolution of identity,
\begin{equation}
-\hat{v}^B_{\rm tot} \Ket{n} + \sum_{j\in B}^{\rm Occ} \hat{v}^B_{nj} \Ket{j}
 \cong \sum_{\varepsilon\in B}^{\rm RI} H_{n\varepsilon}^B \Ket{\varepsilon} \;,
\end{equation}
leads to a much simpler formula
that requires only OEI's and ERI's.
Therefore, by combining
equations~\eqref{e:v-oep-group-1-prefinal}, \eqref{e:v-compl-b-coeff} and \eqref{e:gdf-compl.v}
one arrives to
\begin{equation} \label{e:v-oep-group-1-final}
 V^B_{n\xi} = \sum_{\eta\in B}^{\rm DF} 
          \sum_{\varepsilon\in B}^{\rm RI}
         \left[ {\bf R}^{-1} \right]_{\xi\eta}
         R_{\eta \varepsilon} 
         H_{n\varepsilon}^B \;,
\end{equation}
where
\begin{equation} \label{e:v-oep-group-1-final:H-matrix}
 H_{n\varepsilon}^B = \sum_{\zeta\in B}^{\rm RI} \left[ {\bf S}^{-1} \right]_{\varepsilon\zeta}
   a_\zeta^{(n)}
\end{equation}
and
\begin{align} \label{e:v-oep-group-1-final:a-vector}
 a^{(n)}_\zeta &= -\tBraKet{\zeta}{\hat{v}^B_{\rm tot}}{n}
      + \sum_{j\in B}^{\rm Occ} \tBraKet{\zeta}{\hat{v}^B_{nj}}{j} \nonumber \\
 &=-\sum_{y\in B} W^{(y)}_{\zeta n} 
  + \sum_{j\in B}^{\rm Occ} 
  \left\{
   2\BraKet{\zeta n}{jj} - \BraKet{\zeta j}{nj} 
  \right\} \;.
\end{align}
Note that all the calculations that are required to obtain $V^B_{n\xi}$ are performed
solely on the densities and basis sets associated with the unperturbed molecule $B$.
Therefore, $V^B_{n\xi}$ can be considered as effective fragment parameters
used to compute the first two terms of Eq.~\eqref{e:ct-murell-etal.vin} by
\begin{equation} \label{e:ct.group-i.final}
        -\tBraKet{i}{\hat{v}^B_{\rm tot} }{n} 
      - \sum_{j\in B}^{\rm Occ} \BraKet{nj}{ij} 
       = \sum_{\eta\in B}^{\rm RI} V^B_{n\eta} S_{\eta i} \;,
\end{equation}
which is a great simplification over the original form of group (i)
because only the overlap integrals between the $i$th MO on molecule $A$
and $\eta$th auxiliary orbital on molecule $B$ need to be evaluated. % on the fly.
Note that the only approximation made so far was the application of density fitting
and resolution of identity. If the auxiliary and intermediate
basis sets are sufficiently large, the errors
due to this approximation can be minimal and negligible in principle.

\paragraph{Group (ii).}
The term belonging to this group can be considered as a sum of interaction
energies between the total charge density distribution of molecule $B$
and the partial density $\rho_{ik}({\bf r})$ of molecule $A$,
weighted by the overlap integrals $S_{nk}$. Using the distributed multipole 
expansion based on the charge centroids of the localized molecular orbitals (LMO's),
${\bf r}_i = \tBraKet{i}{\hat{\bf r}}{i}$ with $\chi_i({\bf r})$ being localized,
this group can be approximated by
\begin{equation} \label{e:ct.group-2}
       \sum_{k\in A}^{\rm Occ} S_{nk}  \tBraKet{k}{\hat{v}^B_{\rm tot} }{i} 
 \approx  S_{ni} \rho_{ii}^A \odot \rho^B_{\rm tot} \;.
\end{equation}
Here, it was assumed that
$\lvert\rho_{ik} ({\bf r}) \rvert \ll \rho_{ii} ({\bf r}) $ 
for $i\ne k$ in most locations in the case of LMO's,
which allows one to conjecture that
\begin{equation} \label{e:ct.group-2-d}
 \rho_{ki}^A \odot \rho^B_{\rm tot} \approx \delta_{ik} \rho_{ii}^A \odot \rho^B_{\rm tot} \;.
\end{equation}
Now, the DMTP expansion 
of the interaction energy in the right hand 
side of Eq.~\eqref{e:ct.group-2-d} can be given by
\begin{equation}
 \rho_{ii}^A \odot \rho^B_{\rm tot} 
 \approx 
 q_{i} 
 \left[
  %\sum_{i\in A}^{\rm LC} 
 \sum_{y\in B}^{\rm At}
  \frac{Z_y}{\vert {\bf r}_y - {\bf r}_i \vert } 
 +
 2\sum_{j\in B}^{\rm Occ}
  \frac{q_j}{\vert {\bf r}_j - {\bf r}_i \vert } 
 \right] 
% + \text{ quadrupole terms } 
\;,
\end{equation}
because the distributed charges
$q_i = -1$ whereas the distributed dipole moments 
centered at their
respective LMO charge centroids vanish.\cite{Etchebest.Lavery.Pullman.TheorChimActa.1982}
This means that Eq.~\eqref{e:ct.group-2} can be finally given as follows:
\begin{equation} \label{e:ct.group-ii.final}
       \sum_{k\in A}^{\rm Occ} S_{nk} \tBraKet{k}{\hat{v}^B_{\rm tot} }{i} 
 \approx - S_{ni}  \left[
  %\sum_{i\in A}^{\rm LC} 
 \sum_{y\in B}^{\rm At}
  \frac{Z_y}{ r_{yi} } 
 -
 \sum_{j\in B}^{\rm Occ}
  \frac{2}{r_{ji}} 
 \right] \;.
\end{equation}
Therefore, only overlap integrals and relative distances between
atomic and LMO centroid positions are needed, which leads to a great reduction of the
calculation cost,
as compared either to the original expression or to the multipole expansion (left\hyp{} and right\hyp{}hand sides
of Eq.~\eqref{e:ct.group-2}, respectively).
We shall refer to this approximation as to
the localized overlap approximation (LOA) resulting
in similar expressions to the ones obtained by Jensen and Gordon
in their exchange\hyp{}repulsion interaction energy EFP2 model
(see Eq.~(39) in Ref.\cite{Jensen.Gordon.MolPhys.1996}). Note that, to make this approximation valid,
occupied molecular orbitals need to be spatially localized.

\paragraph{Group (iii).}
The terms with the overlap integrals involving the occupied MO on $A$
can be combined into a single summation term, i.e.,
\begin{multline} \label{e:ct.group-iii}
        \sum_{j\in B}^{\rm Occ} S_{ij} \tBraKet{j}{\hat{v}^A_{i} }{n}  
     + \sum_{k\in A}^{\rm Occ} \sum_{j\in B}^{\rm Occ}  
        S_{kj} \left( 1 - \delta_{ik} \right)
        \BraKet{nj}{ik} \\ = 
 \sum_{k\in A}^{\rm Occ} 
 \sum_{j\in B}^{\rm Occ}
 S_{kj} 
 \tBraKet{j}{
 \underbrace{
 \left\{ 
  \delta_{ik} \left( \hat{v}^A_{k} + \hat{v}^A_{ik} \right)   - \hat{v}^A_{ik}
 \right\} 
  }_{\hat{v}^{A,{\rm eff}}_{ik}}
 }{n}  \\ \cong 
  \sum_{k\in A}^{\rm Occ} 
 \sum_{j\in B}^{\rm Occ}
 S_{kj} 
 \rho_{nj}^B \odot \rho^{A,{\rm eff}}_{ik} \;,
\end{multline}
where the effective potential ${v}^{A,{\rm eff}}_{ik}$ (with the associated 
effective density $\rho^{A,{\rm eff}}_{ik}$) 
is defined by
\begin{equation} \label{e:ct.group-iii.oep}
 {v}^{A,{\rm eff}}_{ik}({\bf r}) \equiv
 \delta_{ik} 
 \left[
  v^A_{\rm tot} ({\bf r}) - 2 v^A_{kk} ({\bf r}) + v^A_{ik} ({\bf r})
 \right] 
  - v^A_{ik} ({\bf r}) \;.
\end{equation}
In order to include the $\rho^{B}_{nj}$ density, it is approximately represented here by a set of effective 
cumulative
atomic charges $\{ q^{B,(nj)}_{y} \}$ associated with the effective one\hyp{}particle density matrix
\begin{equation} \label{e:oed-group-iii}
 P^{B,(nj)}_{\beta\delta} = C_{\beta n} C_{\delta j} \;.
\end{equation}
In this work, the effective charges were defined via the Mulliken method as
discussed in Appendix~\ref{e:a1-v.oep-camm} with $\lambda=0$. 
By applying the LOA in LMO picture for the $\rho_{ik}^A$ density
the effective potential from Eq.~\eqref{e:ct.group-iii.oep} 
simplifies to
\begin{equation} \label{e:ct.group-iii.oep.simplified}
 {v}^{A,{\rm eff}}_{ik}({\bf r}) \approx
 \delta_{ik} \left\{
 {v}^A_{\rm tot} ({\bf r})- 2{v}^A_{kk} ({\bf r}) \right\} \;,
\end{equation}
which
leads to 
\begin{multline} \label{e:ct.group-iii.final.b}
         \sum_{j\in B}^{\rm Occ} S_{ij} \tBraKet{j}{\hat{v}^A_{\rm tot} }{n}  
     + \sum_{k\in A}^{\rm Occ} \sum_{j\in B}^{\rm Occ}  
        S_{kj}
        \BraKet{nj}{ik}  \\
 \approx  
 \sum_{j\in B}^{\rm Occ} S_{ij}
 \sum_{y\in B}^{\rm At} 
 q^{B,(nj)}_{y} 
 \left[ 
   \sum_{x\in A}^{\rm At}
   \frac{Z_x}{r_{xy}}
  + \frac{2}{r_{iy}}
  - \sum_{k\in A}^{\rm Occ}
    \frac{2}{r_{ky}} 
 \right]
 \;.
\end{multline}

\paragraph{Final OEP-based forms of the coupling constant.}
Gathering the results from Eqs.~\eqref{e:ct.group-i.final},
\eqref{e:ct.group-ii.final} and \eqref{e:ct.group-iii.final.b} the coupling constant
can be given as follows
\begin{equation} \label{e:ct-murell-etal.vin.oep}
U_{in}^{A\rightarrow B} 
       \approx 
 G_{1;in}^{A\rightarrow B} 
+G_{2;in}^{A\rightarrow B} 
+G_{3;in}^{A\rightarrow B} \;,
\end{equation}
where
\begin{subequations} \label{e:ct-murrell-etal.vin.oep-gterms}
\begin{align}
 G_{1;in}^{A\rightarrow B} &\equiv \sum_{\eta\in B}^{\rm RI} V^B_{n\eta} S_{\eta i} \;, \\
 G_{2;in}^{A\rightarrow B} &\equiv -S_{ni} u_i^{BA} \;, \\
 G_{3;in}^{A\rightarrow B} &\equiv
  \sum_{j   \in B}^{\rm Occ} S_{ij}
  \sum_{y   \in B}^{\rm At} q_y^{B,(nj)} w_{yi}^{BA}
\;.
\end{align}
\end{subequations}
In the above, the auxiliary variables are
\begin{subequations} \label{e:ct-murell-etal.vin.oep.aux-var}
\begin{align} \label{e:ct-murell-etal.vin.oep.aux-var.u} 
 u_i^{BA} &\equiv    
 \sum_{y\in B}^{\rm At}
  \frac{Z_y}{ r_{yi} } 
 -
 \sum_{j\in B}^{\rm Occ}
  \frac{2}{r_{ji}} 
                \;, \\ 
 w_{yi}^{BA} &\equiv 
   \sum_{x\in A}^{\rm At}
   \frac{Z_x}{r_{xy}}
  + \frac{2}{r_{iy}}
  - \sum_{k\in A}^{\rm Occ}
    \frac{2}{r_{ky}} 
                \;.
  \label{e:ct-murell-etal.vin.oep.aux-var.w}
\end{align}
\end{subequations}
To compute the interaction energy due to CT from $A$ to $B$ by using the above LOA\hyp{}based
approximations, coupling elements need to be transformed from localized to canonical MO basis, i.e.,
\begin{equation} \label{e:ect-oep.prefinal}
 E^{A\rightarrow B} \approx 
 2 
 \sum_{i\in A}^{\rm CMO}
 \sum_{n\in B}^{\rm Vir}
 \frac{
 \lvert
   \sum_{i'\in A}^{\rm LMO} L_{ii'}^A
   U_{i'n}^{A\rightarrow B}
 \rvert^2 }{\varepsilon_i - \varepsilon_n} \;,
\end{equation}
where ${\bf L}^A$ is the CMO\hyp{}LMO transformation matrix for occupied orbitals of $A$. 
Note however, that the effective potential parameters from the density fitted term of group (i) 
do not involve any occupied orbitals.
Therefore, to save computational costs, only groups (ii) and (iii) need to be transformed
whereas for group (i) the overlap integrals from Eq.~\eqref{e:ct.group-i.final} 
can be already evaluated in CMO basis.
This leads to the coupling constant of the following form,
\begin{equation} \label{e:ect-oep.final-vin}
   V_{in}^{A\rightarrow B} \approx 
   G_{1,in}^{A\rightarrow B} \\
  +
   \sum_{i'\in A}^{\rm LMO} L_{ii'}^A
   \left\{
   G_{2,i'n}^{A\rightarrow B}
  +G_{3,i'n}^{A\rightarrow B}
  \right\} \;,
\end{equation}
where the subscripts $G_n$ ($n$ = 1, 2, 3) denote a particular group of 
terms from Eqs.~\eqref{e:ct.group-i.final}, \eqref{e:ct.group-ii.final} 
and \eqref{e:ct.group-iii.final.b}.
Thus, the final working formula for the interaction energy due to CT from $A$ to $B$ reads as
\begin{multline} \label{e:ect-oep.final}
 E^{A\rightarrow B} \approx
 2 
 \sum_{i\in A}^{\rm CMO}
 \sum_{n\in B}^{\rm Vir}
 \frac{1}{\varepsilon_i - \varepsilon_n} \times 
 \Bigg(
   G_{1,in}^{A\rightarrow B} \\
  +
   \sum_{i'\in A}^{\rm LMO} L_{ii'}^A
   \left\{
   G_{2,i'n}^{A\rightarrow B}
  +G_{3,i'n}^{A\rightarrow B}
  \right\}
 \Bigg)^2 \;.
\end{multline}
The total CT energy is given by the sum of the above contribution and the twin contribution
due to CT from molecule $B$ to $A$ according to Eq.~\eqref{e:ct-hf0}.

\section{\label{s:4.calculations}Calculation Details}

Four complexes: 
(i) (H$_2$O)$_2$, 
(ii) H$_2$O--CH$_3$OH, 
(iii) H$_2$O--NH$_4^+$ and 
(iv) NO$_3^-$--NH$_4^+$,
were chosen as model systems to analyse the asymptotic dependence 
of CT/HF0 energy. 
The reference (zero\hyp{}displacement) geometries
were obtained by
performing energy\hyp{}optimizations at the HF/6-31+G(d,p) level,
as implemented in 
the {\sc Gaussian16} quantum chemistry program package.\cite{Gaussian16}
Subsequently, 30 displaced geometries for each model complex
were obtained by translating one of the monomers along the vector 
co\hyp{}linear with the H-bond or N--N distance in the case of ammonium nitrate.
The reference structures as well as the translation vectors are
indicated in the insets of Figure~\ref{f:fig-1} and S1.
To perform statistical error analysis, structural databases
of bi\hyp{}molecular complexes in the non\hyp{}covalent
interactions database NCB31 developed by the Truhlar's 
group,\cite{Zhao.Schultz.Truhlar.JCTC.2006,
Zhao.Truhlar.JCTC.2005,Zhao.Schultz.Truhlar.JCTC.2006,Zhao.Schultz.Truhlar.JCP.2005}
as well as the BioFragment Database subset BBI for backbone\hyp{}backbone
interactions in proteins,\cite{Burns.Faver.Zheng.Marshall.Smith.Vanommeslaeghe.MacKerell.Merz.Sherrill.JCP.2017} 
as implemented in the {\sc Psi4}
program,\cite{Psi4.JCTC.2017}
were utilized.
In particular, the subsets from the NCB31 database were separately considered:
the HB6/04 hydrogen bonding database,\cite{Zhao.Truhlar.JCTC.2005,Zhao.Schultz.Truhlar.JCTC.2006,Zhao.Schultz.Truhlar.JCP.2005}
the DI6/04 dipole interaction database,\cite{Zhao.Truhlar.JCTC.2005,Zhao.Schultz.Truhlar.JCTC.2006,Zhao.Schultz.Truhlar.JCP.2005}
the CT7/04 charge-transfer complex database,\cite{Zhao.Truhlar.JCTC.2005,Zhao.Schultz.Truhlar.JCTC.2006,Zhao.Schultz.Truhlar.JCP.2005}
the WI7/05 weak interaction database,\cite{Zhao.Schultz.Truhlar.JCTC.2006,Zhao.Schultz.Truhlar.JCP.2005,Zhao.Truhlar.JPCA.2005}
and 
the PPS5/05 $\pi$-$\pi$ stacking database.\cite{Zhao.Schultz.Truhlar.JCTC.2006,Zhao.Schultz.Truhlar.JCP.2005,Zhao.Truhlar.JPCA.2005}

The benchmark CT/HF0 energy was estimated
from the following assumption:
\begin{equation}\label{e:ct-pthf-ref}
 E^{\rm CT}_{\rm HF0} \cong  E^{\rm Pol}_{\rm DDS/HF} 
 -  E^{\rm Ind} \;,
\end{equation}
where $E^{\rm Pol}_{\rm DDS/HF}$ is the Hartree\hyp{}Fock polarization energy
defined
according to the density decomposition scheme (here referred as to the DDS) 
developed by Mandado and Hermida\hyp{}Ram{\'o}n,\cite{Mandado.Hermida-Ramon.JCTC.2011} 
whereas $E^{\rm Ind}$ term is the 
approximate pure induction energy due to the polarizability effects.\cite{Stone.TheTheoryOfIntermolecularForces.1996}
The DDS polarization energy encompasses all the charge delocalization
effects that are not associated with the Pauli exclusion principle
in the first order, and is computed by
\begin{equation}\label{e:de-pol-dds}
 E^{\rm Pol}_{\rm DDS/HF} \equiv \Delta E^{\rm HF} 
 - \left(
   E^{\rm Coul}_{\rm DDS/HF}
 + E^{\rm Exch}_{\rm DDS/HF}
 + E^{\rm Rep}_{\rm DDS/HF} 
   \right)\;.
\end{equation}
In the above, $\Delta E^{\rm HF}$ is the total HF interaction
energy,
$E^{\rm Coul}_{\rm DDS/HF}$ is the pure Coulombic interaction energy
between the unperturbed (non\hyp{}interacting) charge densities of the monomers in the complex,
and
$E^{\rm Exch}_{\rm DDS/HF}$ and $E^{\rm Rep}_{\rm DDS/HF}$
are the exchange and the Pauli\hyp{}repulsion interaction energies, respectively.
At the HF level of theory, the sum of the latter two terms,
i.e., the exchange\hyp{}repulsion energy, $E^{\rm Ex-Rep}_{\rm DDS/HF}$,
is in fact
equivalent to the exchange\hyp{}repulsion energy 
in the intermolecular perturbation theory with exchange of Hayes and Stone.\cite{Hayes.Stone.MolPhys.1984} 
It was also shown by 
Mandado and Hermida\hyp{}Ram{\'o}n,\cite{Mandado.Hermida-Ramon.JCTC.2011}
that the Coulombic and exchange\hyp{}repulsion term
in the DDS method reproduce the first\hyp{}order Coulombic and exchange\hyp{}repulsion
interaction energy in SAPT\cite{Jeziorski.Moszynski.Szalewicz.ChemRev.1994}
with accuracy below 0.3 kcal/mole.
Therefore, Eq.~\eqref{e:ct-pthf-ref} is believed to be sufficient
for evaluating the CT/HF0 energy, because there is no dispersion
contribution to the DDS polarization energy at the HF level.

The pure induction energy was estimated by using the distributed
dipole\hyp{}dipole polarizability approximation according to the computational
methodology of Li et al.,\cite{Li.Netzloff.Gordon.JCP.2006}
in which
\begin{equation}\label{e:ind-ref}
 E^{\rm Ind} \approx - \frac{1}{2} {\bf F}^{\rm T} \cdot {\BM \Delta}^{-1} \cdot {\bf F} \;.
\end{equation}
In the above equation, ${\bf F} = \{ \ldots, {\bf F}({\bf r}_a), \ldots\}$
is a super\hyp{}vector of electric fields at the $a$th distributed site
due to the surrounding unperturbed molecular charge densities in the complex
(i.e., the charge\hyp{}density of the molecule that contains $a$th center
is not included in ${\bf F}({\bf r}_a)$).
The matrix elements of ${\BM \Delta}$ are defined as
\begin{equation} \label{e:ind-ref.d-matrix}
 {\BM \Delta}_{ab} = 
\left\{\begin{matrix}
{\BM\upalpha}_a^{-1} &\text{if $a=b$}\qquad\qquad\qquad\qquad\qquad\qquad \\ 
0                    &\text{if $a\ne b \land a,b \in$ same molecule}\\ 
-{\bf T}_{ab}        &\text{if $a\ne b \land a,b \notin$ same molecule}&
\end{matrix}\right.
\end{equation}
and ${\bf T}_{ab}$ is the dipole\hyp{}dipole interaction tensor
given by
\begin{equation} \label{e:ind-ref.t-matrix}
 {\bf T}_{ab} \equiv \frac{3}{r_{ab}^5} {\bf r}_{ab} \otimes {\bf r}_{ab}
  - \frac{1}{r_{ab}^3} {\bf 1} \;,
\end{equation}
with ${\bf r}_{ab} \equiv {\bf r}_{a} - {\bf r}_{b}$ being the
relative distance between the LMO centroids.
The electric fields at the LMO centroids were computed from
the unperturbed HF charge\hyp{}density distributions
\begin{equation} \label{e:ind-ref.efield}
 {\bf F}({\bf r}_a) = \sum_x^{\rm At} \frac{Z_x}{r_{ax}^3} {\bf r}_{ax} 
  + 2 \sum_{\alpha\gamma}^{\rm AO} D_{\alpha\gamma} 
%    \tBraKet{\gamma}{\frac{{\bf r}_a - {\bf r}}{\vert {\bf r}_a - {\bf r} \vert^3} }{\alpha}_{\bf r} \;.
  \left< \gamma \left| 
    \frac{{\bf r}_a - {\bf r}}{\vert {\bf r}_a - {\bf r} \vert^3}
               \right| \alpha \right>_{\bf r} \;,
\end{equation}
in which ${\bf D}$ is the one\hyp{}particle density matrix.
The distributed anisotropic dipole\hyp{}dipole
polarizabilities ${\BM{\upalpha}}_a$ centered at the LMO centroids
were computed by utilizing the coupled\hyp{}perturbed Hartree\hyp{}Fock
method.\cite{McWeeny.RevModPhys.1960,Dodds.McWeeney.Sadlej.MolPhys.1977} 
It was found here that accuracy of the LOA 
from Eqs.~\eqref{e:ct.group-ii.final} and \eqref{e:ct.group-iii.final.b}
is usually slightly better 
when the Boys method\cite{Boys.RevModPhys.1960} is used, as compared
to the Pipek\hyp{}Mezey method\cite{Pipek.Mezey.JCP.1989}. Henceforth, the former
method was used for molecular orbital localization throughout all the production calculations.

All the models that were used to test the theory presented in this work,
i.e., the OL, EFP2 and OEP methods, as well as the EDF and the benchmark CT/HF0 and DDS methods,
were implemented in our in\hyp{}house plugin to {\sc Psi4} quantum chemistry program.\cite{Psi4.JCTC.2017}
DDS method was implemented in the dimer\hyp{}centered basis set\cite{Chalasinski.Gutowski.MolPhys.1985}
in order to eliminate the BSSE and properly inlcude the CT processes.\cite{Stone.Misquitta.CPL.2009}
Estimations of the $E^{\rm Coul}_{\rm DDS/HF}$, $E^{\rm Ex-Rep}_{\rm DDS/HF}$,
$E^{\rm Ind}$ and $E^{\rm CT}_{\rm EFP2}$ interaction energy components 
obtained by using the computer 
code developed for this work were
found to be equivalent with the EFP2 interaction energy components
obtained by using the {\sc GAMESS US} package\cite{GAMESS.JCC.1993} 
(see Supplementary Information, Figure~S2).
For the CT EFP2 component,
potential energy integrals
from Eq.~\eqref{e:coulomb-energy:dmtp-oei} 
were calculated with the
CAMM up to quadrupoles
(distributed centers are atoms), 
instead of the DMA 
(distributed centers are atoms and mid\hyp{}bond points) 
as implemented in most
of quantum chemistry programs. The choice of CAMM versus DMA
was due to convenience of implementation and, because
the quantitative accuracy of our EFP2 CT energy code is comparable to
the EFP2 code of {\sc GAMESS US}, using only atomic distribution centers
does not affect the interpretation
of results in Section~\ref{s:5.results}.

\section{\label{s:5.results}Results and Discussion}

\subsection{\label{ss:5.1.accuracy}Accuracy of OEP-based model}

\begin{figure}[h]
\includegraphics[width=0.44\textwidth]{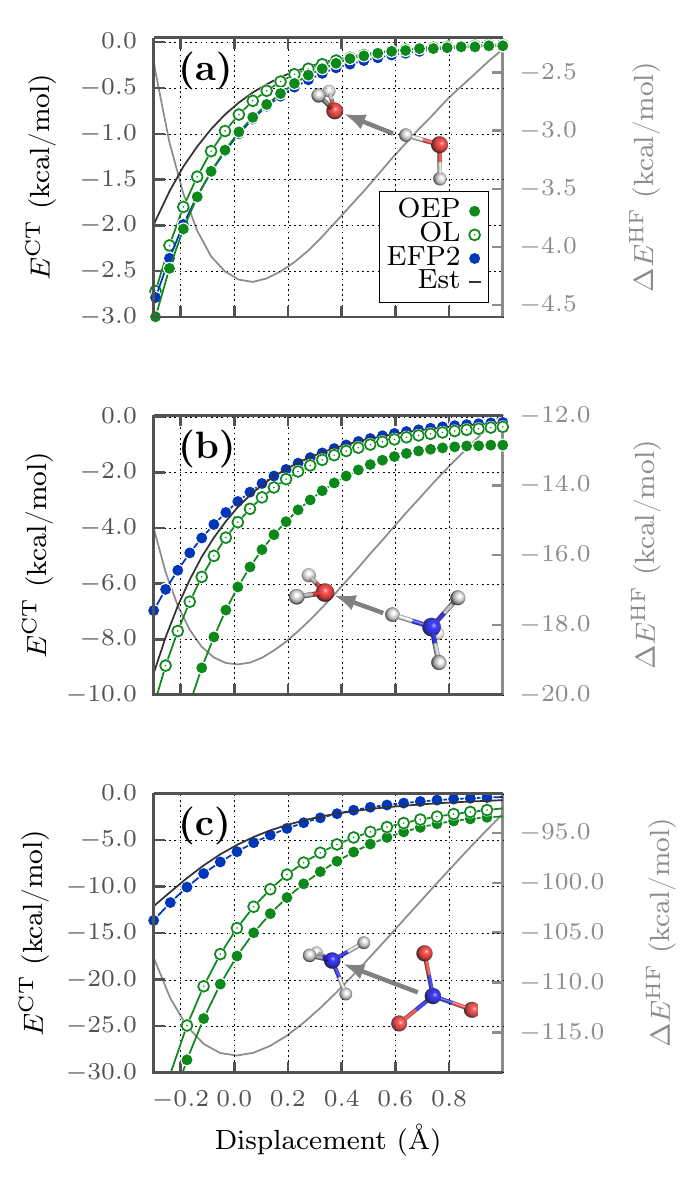}
\caption{\label{f:fig-1} {\bf Asymptotic dependence of the charge transfer energy
in the CT/HF0 formulation for selected bi\hyp{}molecular complexes.} 
(a) water dimer, 
(b) water\hyp{}ammonium complex, and 
(c) ammonium\hyp{}nitrate complex,
were one molecule has been translated
along the vector specified in the insets relative to initial geometry,
optimized at HF/6-31+G(d,p) level.
The 
total interaction energy
is also shown for comparison in light grey color in this figure.
Interaction energies were obtained at the HF/6-311++G(2df,2pd) level of theory
with the single GDF scheme and aug-cc-pVQZ-jkfit auxiliary basis set.
} 
\end{figure}
The asymptotic dependence of the CT energy
is
shown in Figure~\ref{f:fig-1} and Figure~S1 for the methanol\hyp{}water system. 
Overall, OEP technique of effective
elimination of ERI's sketched in Eq.~\eqref{e:ft-reduction} 
is sufficiently accurate
for water dimer and ammonium nitrate (Figures~\ref{f:fig-1}(a) and (c)),
and methanol\hyp{}water system (Figure~S1) in all distances studied,
reproducing the reference Otto\hyp{}Ladik CT energy, with errors not
exceeding 10\% relative to OL estimates (compare open and filled green circles).
However, in the case of water\hyp{}ammonium cation complex the OEP
method overestimates the CT energy by around 50\%, even for large
intermonomer separations, where the profile exhibits
unphysical long\hyp{}range tail, visible in Figure~\ref{f:fig-1}(b). 
Careful inspection
(see Figure~S3)
revealed that increasing the size of the 
auxiliary basis set from aug-cc-pVDZ-jkfit to aug-cc-pVQZ-jkfit
slightly improves the performance of the OEP method, however the long\hyp{}range
tail is still appreciable in magnitude even in the aug-cc-pVQZ-jkfit auxiliary basis.

At the moment, it is unclear what causes these errors at large separations. 
The LOA in $G_2$ and $G_3$ may be too drastic approximations
when one of the monomers is charged, and more accurate
DMTP expansion for the effective densities could be required,
also including dipole and higher multipole moments. % and as well as finer distribution of centres.
Interestingly in the case of ammonium nitrate (Figure~\ref{f:fig-1}(c)),
the relative errors between OEP and OL estimates
are not grater than 9\% even though both of the monomers are charged.
On the other hand, OL method itself strongly overestimates the CT/HF0
energies in this system. Therefore, among the four
model complexes, EFP2 and OEP energies
coincide with the CT/HF0 estimates 
only in the case of water dimer and methanol-water complex,
in contrast to the EFP2 model, which gives very good estimates
of the CT/HF0 energy in mostly all cases at all separations.

\begin{figure*}[h]
\includegraphics[width=0.9\textwidth]{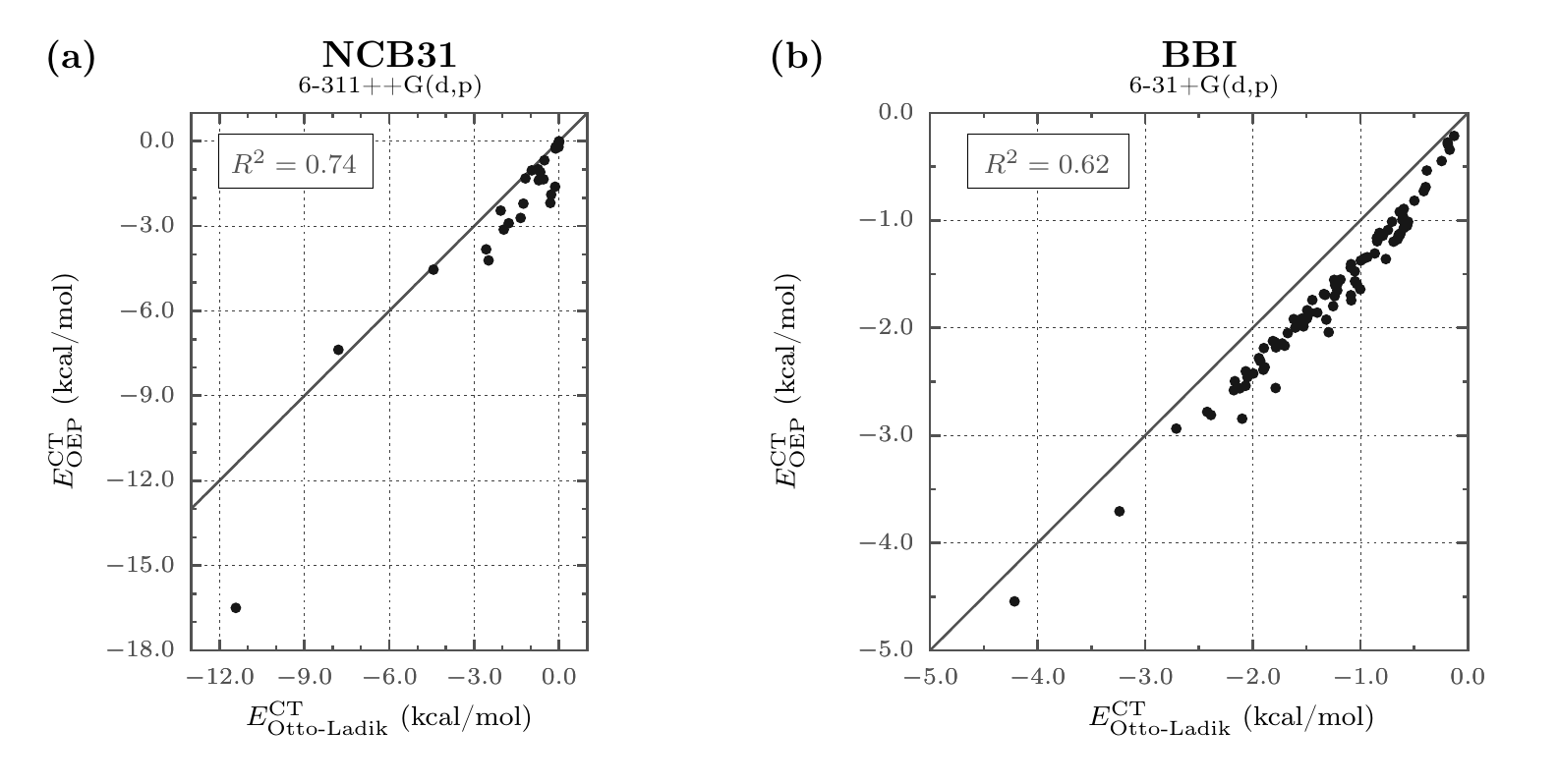}
\caption{\label{f:fig-2} {\bf Accuracy of the ERI elimination technique.}
(a) NCB31 
database,\cite{Zhao.Schultz.Truhlar.JCTC.2006,
Zhao.Truhlar.JCTC.2005,Zhao.Schultz.Truhlar.JCTC.2006,Zhao.Schultz.Truhlar.JCP.2005} 
and
(b) BBI subset\cite{Burns.Faver.Zheng.Marshall.Smith.Vanommeslaeghe.MacKerell.Merz.Sherrill.JCP.2017} 
from the BioFragment Database.
For the OEP calculations, the EDF-1 scheme with the aug-cc-pVDZ-jkfit auxiliary basis set
was used.
} 
\end{figure*}

\begin{figure*}[h]
\includegraphics[width=0.9\textwidth]{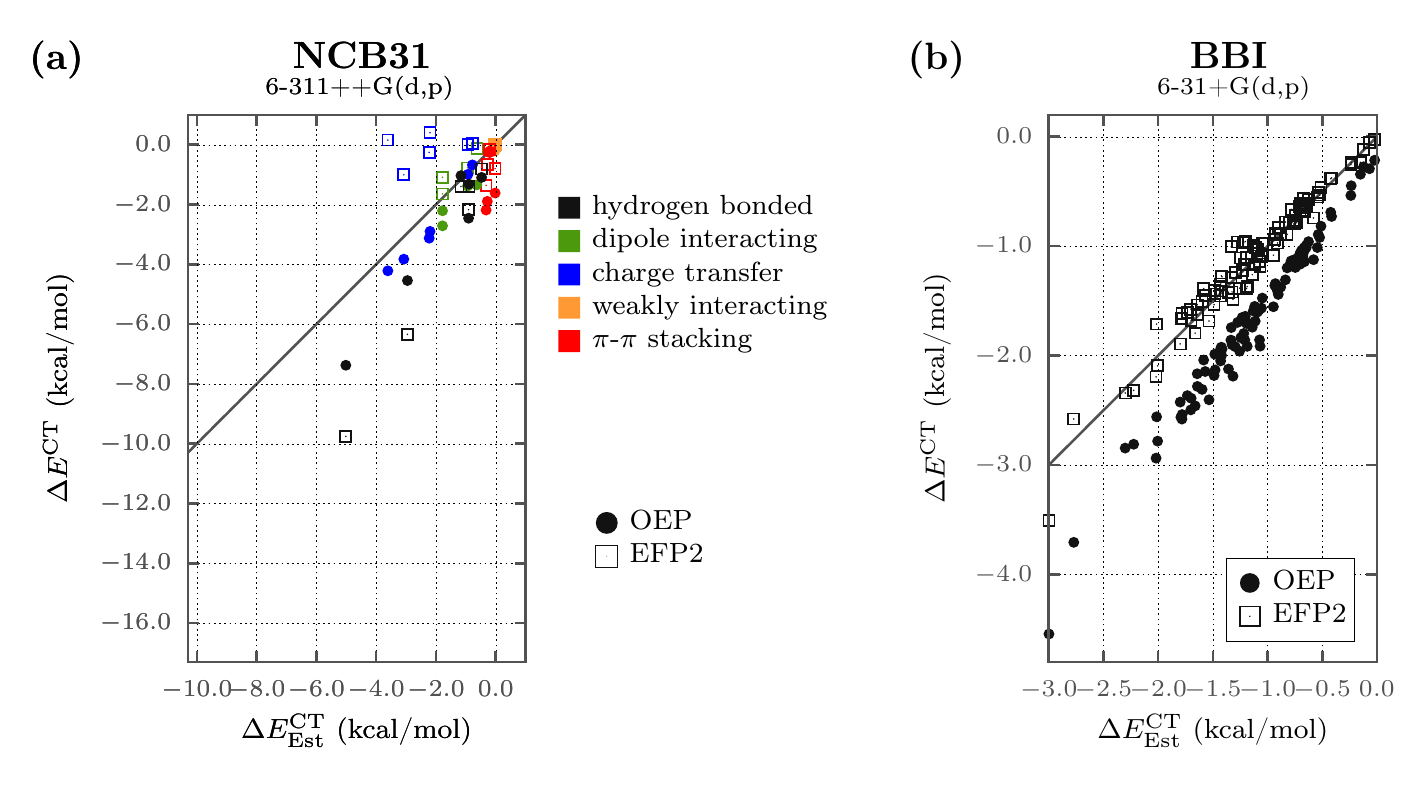}
\caption{\label{f:fig-3} {\bf 
Performance of the OEP and EFP2 methods for the CT/HF0 charge transfer interaction energy
in bi\hyp{}molecular neutral complexes.}
(a) NCB31 set from database
developed by the Truhlar's group\cite{Zhao.Schultz.Truhlar.JCTC.2006,
Zhao.Truhlar.JCTC.2005,Zhao.Schultz.Truhlar.JCTC.2006,Zhao.Schultz.Truhlar.JCP.2005}:
the HB6/04 hydrogen bonding database 
(black),\cite{Zhao.Truhlar.JCTC.2005,Zhao.Schultz.Truhlar.JCTC.2006,Zhao.Schultz.Truhlar.JCP.2005}
the DI6/04 dipole interaction database 
(green),\cite{Zhao.Truhlar.JCTC.2005,Zhao.Schultz.Truhlar.JCTC.2006,Zhao.Schultz.Truhlar.JCP.2005}
the CT7/04 charge-transfer complex database 
(blue),\cite{Zhao.Truhlar.JCTC.2005,Zhao.Schultz.Truhlar.JCTC.2006,Zhao.Schultz.Truhlar.JCP.2005}
the WI7/05 weak interaction database 
(yellow),\cite{Zhao.Schultz.Truhlar.JCTC.2006,Zhao.Schultz.Truhlar.JCP.2005,Zhao.Truhlar.JPCA.2005}
and the PPS5/05 the $\pi$-$\pi$ stacking database 
(red);\cite{Zhao.Schultz.Truhlar.JCTC.2006,Zhao.Schultz.Truhlar.JCP.2005,Zhao.Truhlar.JPCA.2005}
(b) BBI subset\cite{Burns.Faver.Zheng.Marshall.Smith.Vanommeslaeghe.MacKerell.Merz.Sherrill.JCP.2017} 
from the BioFragment Database.
The EDF-1 scheme with the aug-cc-pVDZ-jkfit auxiliary basis set
for the OEP calculations.
} 
\end{figure*}

Analysis of the statistical sets of dimeric complexes
shown in Figures~\ref{f:fig-2}(a) and Figures~\ref{f:fig-3}(a)
suggests that, at least in the case of neutral systems,
ERI elimination technique (compare OL and OEP)
is of acceptable accuracy with the $R^2$ coefficients
being 74\% for the NCB31, and 62\% for the BBI dataset. 
The absolute accuracy of the OEP model, measured with respect to the benchmark
CH/HF0 estimates, is shown in Figures~\ref{f:fig-2}(b) and Figures~\ref{f:fig-3}(b)
and the root mean square errors (RMSE)
are listed in Table~\ref{t:2}. In NCB31 set OEP performs better than EFP2
with
RMSE of 1.69 and 2.39 kcal/mol, respectively. This is especially
the case for charge\hyp{}transfer\hyp{}dominated systems (subset CT7/04),
in which EFP2 erroneously predict vanishing or even small
positive CT energiers (RMSE 4.06), in contrast to OEP method that gives correct
signs in all cases and good correlation with the benchmark.
On the other hand, EFP2 method very accurately reproduces the CT/HF0 energies
in the BBI set (R$^2=95\%$) whereas OEP model tends to significantly overestimate
the values by a factor of 1.55 with RMSE of 0.56 kcal/mol, 
unlike the OL model which is more accurate
(RMSE of 0.22 and R$^2=78\%$). 
{
\renewcommand{\arraystretch}{1.4}
\begin{table}[t]
\caption[Accuracy of approximate CT/HF0 energy methods across wide range of bi\hyp{}molecular complexes]
{{\bf Accuracy of approximate CT/HF0 energy methods across wide range of bi\hyp{}molecular complexes\footnotemark[1]}
}
\label{t:2}
\begin{ruledtabular}
\begin{tabular}{lclcccccccc}
         &&       && \multicolumn{7}{c}{RMSE (kcal/mol) } \\
\cline{4-11}
\multicolumn{3}{c}{Database}
   && OL\footnotemark[2]  
                  && \multicolumn{3}{c}{OEP\footnotemark[3]}  
                  && EFP2\footnotemark[4] \\
\cline{1-3}
\cline{4-5}
\cline{6-9}
\cline{10-11}
NCB31\footnotemark[5] 
      &&  HB6/04  && 1.38 && 1.36 && 0.37 && 2.45 \\ % 
      &&  DI6/04  && 0.35 && 0.66 && 0.22 && 0.44 \\ % 
      &&  CT7/04  && 0.98 && 2.82 && 0.71 && 4.06 \\ % 
      &&  WI7/04  && 0.01 && 0.09 && 0.05 && 0.02 \\ % 
      &&  PPS5/05 && 0.07 && 1.31 && 0.79 && 0.60 \\ % 
      &&  Total   && 0.83 && 1.69 && 0.53 && 2.39 \\ % 
\cline{1-3}
\cline{4-5}
\cline{6-9}
\cline{10-11}

BBI\footnotemark[6]   
      &&  Total
                  && 0.26 && 0.56 && 0.13 && 0.12 \\ %
\end{tabular}
\end{ruledtabular}
\footnotetext[1]{Validated against Eq.~\eqref{e:ct-pthf-ref}.}
\footnotetext[2]{Otto and Ladik's expression from Eq.~(4) in Ref.~\cite{Otto.Ladik.ChemPhys.1975}}
\footnotetext[3]{This work, Eq.~\eqref{e:ect-oep.final-scaled} 
with $C=1.00$ (left column) and $C=1.56$ (right columm).
GDF(1)/aug-cc-pVDZ-jkfit scheme was used for the extended density fitting of group (i) OEP's.}
\footnotetext[4]{Ref.~\cite{Li.Gordon.Jensen.JCP.2006}, Eq.~\eqref{e:u-efp2-ct}.}
\footnotetext[5]{Ref.~\cite{Zhao.Truhlar.JCTC.2005,Zhao.Schultz.Truhlar.JCTC.2006,Zhao.Schultz.Truhlar.JCP.2005,Zhao.Schultz.Truhlar.JCTC.2006}}
\footnotetext[6]{Ref.~\cite{Burns.Faver.Zheng.Marshall.Smith.Vanommeslaeghe.MacKerell.Merz.Sherrill.JCP.2017}}
\end{table}
}

Since the nature of the EFP2 and OEP models is highly approximate,
and the most important goal is to construct efficient and reasonably
accurate \emph{ab initio} force field, one can use the fact that
OEP model almost always overestimates the CT/HF0 energies
with rather good linear correlation,
unlike EFP2 model, which can also strongly underestimate it.
Therefore, the simple semi\hyp{}empirical scaling model is proposed here
for the OEP\hyp{}based estimate, i.e.,
\begin{equation} \label{e:ect-oep.final-scaled}
 E^{\rm CT}_{\rm scaled} \approx C \times E^{\rm CT}_{\rm unscaled}
\end{equation}
where $C$ is an empircial constant and $E^{\rm CT}_{\rm unscaled}$ is given by
Eqs.~\eqref{e:ct-hf0} and Eq.~\eqref{e:ect-oep.final}. Based on
OEP performance from NCB31 and BBI sets, it was estimated
that $C\approx 1.56$. Application of the semi\hyp{}empirical model
from Eq.~\eqref{e:ect-oep.final-scaled} significantly improved
absolute accuracy of the model in all datasets. 
For instance, RMSE was reduced down to 0.13 kcal/mol in the BBI set,
comparable to EFP2 model.
Naturally, the accuracy of the OL model limits the absolute accuracy
of the OEP model, which is manifested in the ammonium nitrate system
in Figure~\ref{f:fig-1}(c) or the problems with LOA. 
In order to improve the performance for ionic systems,
further studies are necessary, that will modify the original formulation
of OL and LOA treatment, 
or use different sets of virtual orbitals\cite{Xu.Gordon.JCP.2013}. Nevertheless, except from the ionic systems,
it is clear that the level of accuracy of the EFP2 and OEP models are rather comparable,
especially after applying the empirical scaling,
across a wide range of various interacting complexes including 
systems with H-boding, dominated by dipole-dipole or dispersion interactions or 
exhibiting substantial CT.

{
\renewcommand{\arraystretch}{1.4}
\begin{table}[t]
\caption[Estimated computational cost of the EFP2 and OEP methods for calculation of CT/HF0 energy]
{{\bf Estimated computational cost of the EFP2 and OEP methods for calculation of CT/HF0 energy\footnotemark[1]}
}
\label{t:oep-costs}
\begin{ruledtabular}
\begin{tabular}{lccc}
Method     && EFP2 & OEP \\
\hline
%Constants  && $\varepsilon_i$ 
%            & $\varepsilon_i$, $\varepsilon_n$ \\
Constant   && \multirow{2}{*}{$\varepsilon_i$} 
            & \multirow{2}{*}{$\varepsilon_i$, $\varepsilon_n$, $L_{ii'}^A$} \\
parameters\footnotemark[2] 
           && & \\
\hline
%\multirow{2}{*}{Rotatable}  
Superimposable  && \multirow{2}{*}{$C_{\alpha i}^A$, $C_{\beta n}^B$, $\{\alpha\}$, $\{\beta\}$  } 
            & $C_{\alpha i}^A$, $C_{\alpha n}^B$, $\{\alpha\}$, $\{\beta\}$, \\
parameters\footnotemark[2] 
                && & $\{\eta\}$, $V_{n\eta}^{{\rm eff},B}$ \\
\hline
\multirow{3}{*}{Calculables\footnotemark[2]} 
           && $S_{ij}$, $S_{nk}$, $S_{nw}$,            &   $S_{ij}$, $S_{\eta i}$, $S_{ni}$, \\
           && $T_{nn}$, $T_{kj}$, $T_{wj}$, $T_{nj}$,  &  $u_i^{BA}$, $w_{yi}^{BA}$\\
           && $U_{in}^{{\rm EF},B}$ , $U_{ik}^{{\rm EF},B}$, $U_{iw}^{{\rm EF},B}$
                                                       &  \\
\hline
\multirow{4}{*}{Cost\footnotemark[3]}    
                         && $sp\left( 2p^2+2op+o^2\right)$ & \\
                         &&$+tp\left( 2p^2+2op+o^2\right)$ & $sop\left( 2p+o+a\right)$ \\ 
                         &&$+vop\left( 3p+o\right)$        & $+op(a+oN+2o)$\\             
                         && $+o^2p$  & \\
\end{tabular}
\end{ruledtabular}
\footnotetext[1]{Based on coupling constant
expressions from Eq.~\eqref{e:u-efp2-ct-v2} and Eq.~\eqref{e:ect-oep.final-vin} 
for EFP2 and OEP
method, respectively.}
\footnotetext[2]{The subscript meaning is as follows: 
primary basis set functions of $A$: $\alpha$;
primary basis set functions of $B$: $\beta$;
auxiliary basis set functions of $B$: $\eta$,
occupied MO's of $A$: $i$, $i'$, $k$;
occupied MO's of $B$: $j$;
virtual MO's of $A$: $w$; 
virtual MO's of $B$: $n$; 
atoms of $B$: $y$. Analysis is based on $E^{A\rightarrow B}$ term.}
\footnotetext[3]{
Numbers of: 
primary basis set functions - $p$;
auxiliary basis set functions - $a$;
occupied MO's - $o$;
atoms - $N$. 
Relative costs: $v$ - multipole potential, $t$ - kinetic energy and $s$ - overlap OEI's.
It was assumed that the number of virtual orbitals is equal to $n$.}
\end{table}
}

\subsection{\label{ss:5.2.cost}Reduction of computational costs}

The utmost goal of this work is to reduce the computational cost of
the CT/HF0 energy evaluation in the calculations involving effective fragment
potentials, no longer making it the bottleneck of the EFP2\hyp{}based interaction
energy calculations. In Table~\ref{t:oep-costs} estimation of the
computational cost of the EFP2 and OEP models is shown.
It is apparent that EFP2 requires 
much more quantities to be computed 
as compared to the OEP method (`Calculables' in the table).
Clearly,
evaluation of the EFP2 CT expression from Eq.~\eqref{e:u-efp2-ct} 
involves quite a number of different types of OEI's. According to our estimations
that assume sequential (two\hyp{}step) two\hyp{}index AO\hyp{}MO transformations of OEI matrices
and large AO basis sets,
the computational cost
is of an order of $2p^3(s+t) + 3vop^2$, where the
$o$ and $p$ denote the number of occupied orbitals and the number of atomic basis functions,
respectively. 
Here, $s$, $t$ and $v$ are the relative costs of evaluation of the
overlap, kinetic energy, and multipole potential OEI's, respectively, with the latter being
most expensive but
necessary to compute ${\bf U}^{\rm EF}$ matrices
from Eq.~\eqref{e:v-efp2-eff}. 
{
\renewcommand{\arraystretch}{1.4}
\begin{table}[t]
\caption[Minimal uncontracted auxiliary basis set optimized for OEP\hyp{}based CT/HF0 calculations]
{{\bf Minimal uncontracted auxiliary basis set optimized for OEP\hyp{}based CT/HF0 calculations\footnotemark[1]}
}
\label{t:3}
\begin{ruledtabular}
\begin{tabular}{lldcd}
 AO's &  & \multicolumn{3}{c}{Exponents (a.u.)}\\
      &  & \text{Methanol}         && \text{Water}           \\
\cline{1-2}
\cline{3-5}
 H    &1s&   7.7369294534   &&   29.5837988322 \\
 C    &1s& 880.3654511045   &&                 \\
      &2s& 123.8276940826   &&                 \\
      &2p&  11.6748155072   &&                 \\
 O    &1s& 902.9798919021   && 1030.5721050297 \\
      &2s& 203.7584408153   &&  142.1022267153 \\
      &2p&  20.2442429692   &&   40.9175702474 \\
\end{tabular}
\end{ruledtabular}
\footnotetext[1]{Fitting performed according to Appendix~\eqref{a:auxiliary-basis} assuming 6-311++G(d,p) 
and aug-cc-pVDZ-jkfit primary and test basis sets, respectively.}
\end{table}
}
\noindent \hspace{-20pt}
On the contrary, 
OEP\hyp{}based expression
from Eq.~\eqref{e:ect-oep.final} requires only overlap OEI's that are the least expensive, 
and has the cost of approximate magnitude of $2sop^2$ for relatively small auxiliary basis sets.
Note also that, among the calculables that are needed in each
OEP\hyp{}based CT energy evaluation, are the auxiliary vectors and matrices 
from Eqs.~\eqref{e:ct-murell-etal.vin.oep.aux-var.u}
and \eqref{e:ct-murell-etal.vin.oep.aux-var.w},
the cost of which is negligible. The amount of effective fragment parameters,
that needs to be superimposed during the calculations by applying rotation
of orbitals and basis functions (`Superimposable parameters')
is rather the same in EFP2 and OEP models. 
This includes the LCAO\hyp{}MO coefficients in canonical basis,
and the primary basis set, with an addition of the auxiliary basis set
for the OEP model. Therefore, the cost of parameter superimposition should
not be significantly larger as in the EFP2 formulation, provided sufficiently 
small auxiliary basis set is used.
For example,
assuming a water dimer system and 6-311++G(d,p) primary and minimal auxiliary basis set
with $s=t=v\approx1$,
the OEP method is predicted to be roughly 12--16 times faster than EFP2 method. In practice,
the parameters $s$, $t$ and $v$ will have larger values, especially $v$.

{
\renewcommand{\arraystretch}{1.4}
\begin{table}[h]
\caption[Minimal auxiliary basis set optimized for OEP\hyp{}based CT/HF0 calculations.]
{{\bf CPU timings in miliseconds of CT/HF0 single point energy calculations
for water-methanol complex at 6-311++G(d,p) primary basis set\footnotemark[1]}
}
\label{t:5}
\begin{ruledtabular}
\begin{tabular}{llccccc}
                      && \multicolumn{2}{c}{Water-Methanol} && \multicolumn{2}{c}{Water dimer} \\
\cline{3-4}
\cline{6-7}
OL                    &&1.48$\times 10^{4}$ 
                                &(-0.81)&& 2.91$\times 10^{3}$     
                                                    &(-0.85) \\  %14716.1 &&  2876.9 added 33 sec and 17 sec for ERI/AO calculation from timer.dat
EFP2                  &&  42.7  &(-1.27)&&   15.6   &(-1.06) \\
OEP/aug-cc-pVQZ-jkfit &&   4.36 &(-1.03)&&    1.64  &(-1.05) \\
OEP/aug-cc-pVTZ-jkfit &&   3.63 &(-1.07)&&    1.35  &(-1.09) \\
OEP/aug-cc-pVDZ-jkfit &&   3.60 &(-1.05)&&    1.26  &(-1.03) \\
OEP/6-311++G(d,p)     &&   2.74 &(-1.37)&&    0.933 &(-1.31) \\
OEP/mini\footnotemark[2]&& 2.25 &(-1.34)&&    0.623 &(-1.13) \\
\end{tabular}
\end{ruledtabular}
\footnotetext[1]{1.2 GHz AMD EPYC\texttrademark{} 7301 16-Core Processor, calculations performed on 1 core. 
CT/HF0 energies are given in parentheses for reference (kcal/mol).
See also the implementation details in Section~\ref{s:4.calculations} and Section~\ref{ss:5.2.cost}.}
\footnotetext[2]{This work, Table~\ref{t:3}.}
\end{table}
}

\begin{figure}[t]
\includegraphics[width=0.46\textwidth]{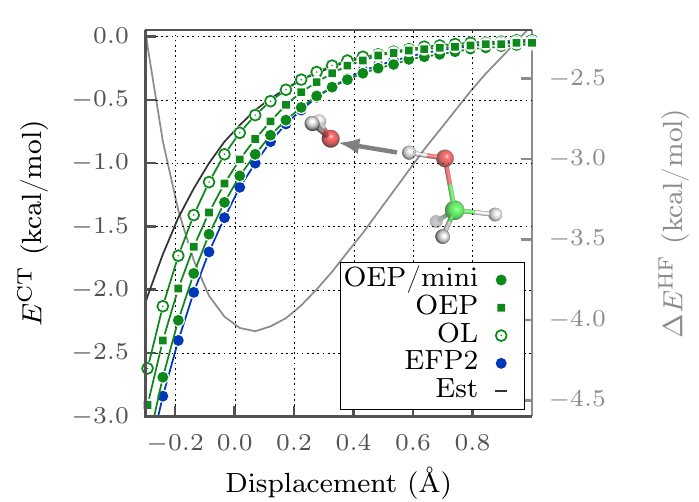}
\caption{\label{f:fig-4} {\bf Asymptotic dependence of the charge transfer energy
in the CT/HF0 formulation for methanol-warer complex with minimal OEP auxiliary basis.} 
Water molecule has been translated by $R$ from the starting geometry
optimized at HF/6-31+G(d,p) level,
along the vector specified in the insets.
The total interaction energy
is also shown for comparison in light grey color in this figure.
Interaction energies were obtained at the HF/6-311++G(d,p) level of theory.
In this figure, `OEP' denotes single GDF scheme with aug-cc-pVQZ-jkfit auxiliary basis set
whereas `OEP/mini` denotes double GDF scheme with the minimal auxiliary basis
given in Table~\ref{t:3}.
} 
\end{figure}

In this work, the minimal uncontracted auxiliary basis sets were optimized for water and methanol
that consist of only $s$ and $p$\hyp{}type functions,
and the orbital exponents are shown in Table~\ref{t:3}
whereas the procedure is outlined in Appendix~\ref{a:auxiliary-basis}.
Unfortunately, we were not able to successfully fit the auxiliary basis sets
for NH$_4^+$ and NO$_3^-$ using current approach
and the basis set optimization falls out of scope of this work.
Nevertheless, we found that in the case of neutral systems minimal basis
sets are already sufficient to reach required accuracy.
In Figure~\ref{f:fig-4} the asymptotic dependence of the CT/HF0 energy
for water\hyp{}methanol system is correctly reproduced by the OEP/mini variant
which
performs comparably well as the OEP/aug-cc-pVQZ-jkfit. 
The total time required for evaluation of the CT/HF0 energies by using EFP2 and OEP
models was measured for this system and also for the water\hyp{}water system in their reference geometries,
and the results are shown in Table~\ref{t:5}. 
Time profiling of the code for OL, EFP2 and OEP methods was performed 
for all the computational operations required for a
single point calculation in a hypothetical sequential run on multiple
geometries. Therefore, calculations of intermonomer 
ERI's in AO basis 
along with their four\hyp{}index transformation to MO basis were included in
time profiling of the OL routine in our computer code. For the EFP2 
and OEP methods, all the calculables were taken into account in the profiling,
i.e., calculations of OEI's in AO basis and their two\hyp{}index transformations
to MO basis. Calculations of canonical orbital 
energies and LCAO\hyp{}MO coefficients through the self\hyp{}consistent field
HF procedure, as well as calculations of the
OEP matrices from Eq.~\eqref{e:v-oep.ct} 
via EDF-2 scheme and atomic effective charges from Eq.~\eqref{e:a1-v.oep-camm.q}
using the OED in Eq.~\eqref{e:oed-group-iii}
were not included because these quantities constitute the effective fragments
by definition.
The above time profiling setting allows one to compare the performances
of the OL, EFP2 and OEP methods as effective fragment parameter methods.

Even for the rather large quadruple\hyp{}$\zeta$ auxiliary basis set,
the OEP\hyp{}based model is roughly 10 times more efficient than evaluating the EFP2 CT energy.
Reducing the size of auxiliary basis set up to minimal basis from Table~\ref{t:3}
scales down the cost further by a factor of $2.0$ in water\hyp{}methanol system,
and $2.6$ in water dimer system, reaching in total roughly 20\hyp{}fold
reduction in CPU time as compared to the EFP2 model, and more than three orders of magnitude
as compared to the OL model. Since the size of such a minimal basis
is very small when compared to the primary basis sets recommended for constructing
the EFP2 parameters, the additional superimposition costs of the auxiliary basis
should be negligible. 
It is also worth mentioning here that, while using the usual 6-311++G(d,p) basis
in a role of auxiliary basis set 
gave already accurate results, the minimal STO-3G basis as auxiliary space resulted
in CT energies of $-8.4$ and $-8.8$ kcal/mol in water\hyp{}methanol and water dimer systems (data not shown),
much lower than the reference values of $-0.81$ and $-0.85$ kcal/mol, respecively. 
Therefore, the basis set optimization
described in Appendix~\ref{a:auxiliary-basis} was necessary for the minimal basis to be applicable
in the CT calculations using OEP technique, with OEP's generated by the EDF-2 scheme
developed in Section~\ref{sss:2.3.1.GDF-2}. 

\section{\label{s:6.conclusions}Summary and a few concluding remarks}

In this work, the one\hyp{}electron effective potential operator technique
of eliminating electron repulsion integrals in the fragment\hyp{}based
theories of intermolecular interactions
was proposed.
It was shown that in general case two types of OEP's
can be defined and worked out either through the density fitting or the distributed
multipole expansion. For the first group of OEP's, the density fitting was extended
and two computational treatments of OEP's were developed: the EDF-1 scheme
for RI\hyp{}type auxiliary basis sets, as well as the EDF-2 scheme, for
essentially all other types of basis sets. 
The OEP technique was then applied to calculating the charge transfer energy
in condensed phases simulations. 
The presented validation of the OEP technique
against the Otto\hyp{}Ladik CT/HF0 model as parent theory showed that in most cases
elimination of ERI's is quantitatively accurate, with few exceptions
where only qualitative accuracy was found. 
Finally, it was concluded that, while the proposed OEP\hyp{}based Otto\hyp{}Ladik
model is of comparable accuracy as the CT/HF0 formulation within the
state\hyp{}of\hyp{}the\hyp{}art EFP2 model, it significantly outperforms the latter
in terms of computational efficiency reaching speedups up to 20 times.
It should be emphasized that there is still more room for optimization of the
OEP technique. First of all, not only group (i) terms, but also
all the summations over the virtual orbitals can be optimized,
e.g., by using the QUAMBO's, following Xu and Gordon.\cite{Xu.Gordon.JCP.2013}
Secondly, effective atomic charges $q_y^{B,(nj)}$ from Eq.~\eqref{e:ct.group-iii.final.b}
are most of the time very small
and could be neglected, leading to further cost reductions (not shown here).
In addition, the possibility of the OEP\hyp{}based formulation of the direct EFP2 CT formulation
should be studied in the future as well.
Therefore, it is believed that the OEP\hyp{}based
model can be incorporated within the EFP2 method, strongly
facilitating
the full (CT\hyp{}including) EFP2 energy calculations and 
molecular dynamics simulations in large systems. However, the current model
needs to be improved by further investigating the ionic
systems in detail and possibly revising the LOA and EDF approximations.
Nevertheless,
it is anticipated here that the OEP method of ERI elimination can be used in 
virtually any other 
\emph{ab initio} fragment\hyp{}based
approaches for condensed\hyp{}phase simulations, where ERI's pose the
computational challenge when confronted with the size of the system.

\begin{acknowledgments}
This project is carried out under POLONEZ programme which has received funding from the European Union's
Horizon~2020 research and innovation programme under the Marie Sk{\l}odowska-Curie grant agreement 
No.~665778. This project is funded by National Science Centre, Poland 
(grant~no. 2016/23/P/ST4/01720) within the POLONEZ 3 fellowship.
Wroclaw Centre for Networking and Supercomputing (WCSS) is acknowledged for
computational resources.
We cordially thank Professor Marcos Mandado from University of Vigo in Spain
for providing us with his benchmark calculations to validate our implementation
of the DDS method.
\end{acknowledgments}

\appendix

\section{Optimized Auxiliary Basis Sets for OEP Applications\label{a:auxiliary-basis}}

To fit the auxiliary DF basis for the treatment of the overlap\hyp{}like
OEP matrix elements with the operator $\hat{v}_{\rm eff}$, 
the following objective function is minimized
\begin{equation} \label{e:a1-obj}
 Z[\{\xi\}] = \sum_{\alpha i} \left[ 
     \tBraKet{\alpha}{\hat{v}_{\rm eff}}{i} - 
     \sum_{\xi}^{\rm DF} V_{\xi i} S_{\alpha \xi} 
    \right]^2 \;,
\end{equation}
where $\{\xi\}$ is the auxiliary basis set to optimize, whereas $\{\alpha\}$
is the `test' basis set used to probe the accuracy of the density fitting.
The forms of the %Pauli and 
charge\hyp{}transfer DF matrices $\tBraKet{\alpha}{\hat{v}_{\rm eff}}{i}$
can be directly derived 
from Eq.~\eqref{e:v-oep.ct}
%from Eqs.~\eqref{e:v-oep.rep} and \eqref{e:v-oep.ct}, respectively, 
by expanding the MO's in terms of the AO's. The working formula
is %given below:
\begin{equation} \label{e:a1-v.oep}
   \tBraKet{\alpha}{\hat{v}_{\rm eff}^{{\rm G}_1}}{n} 
     = -\sum_{x}^{\rm At} W_{\alpha n}^{(x)}   
       + \sum_{\beta\gamma\delta} 
           \left\{
             2 C_{\beta n} D_{\gamma\delta} - C_{\gamma n} D_{\beta \delta}
           \right\}
           \BraKet{\alpha\beta}{\gamma\delta}
%    \;.
\end{equation}
for the OEP operator defined by 
\begin{equation}
 \sum_{\eta\in B}^{\rm DF} \hat{v}_{\rm eff}^{{\rm G}_1} \Ket{\eta} \equiv 
 \sum_{\eta\in B}^{\rm DF} V^B_{n\eta} \Ket{\eta} \;.
\end{equation}
In Eq.~\eqref{e:a1-v.oep}, ${\bf C}$ is the LCAO\hyp{}MO matrix whereas ${\bf D}$
is the one\hyp{}particle density matrix in AO basis.

\section{Coulomb-like OEP via Cumulative Atomic Multipole Moments\label{a:oep-camm}}

Overlap\hyp{}like OEP's are always associated with the effective one\hyp{}particle density
(or bond\hyp{}order) 
matrices and the $\lambda$ parameter from Eq.~\eqref{e:oep-operator} being either 1 or 0. 
Therefore, one can use this fact to compute the effective distributed multipole
moments. In this work, the cumulative atomic multipole moments (CAMM)
of Sokalski and Poirier with Mulliken partitioning of the AO space
were chosen because of their simplicity.
One of examples
of such OEP\hyp{}based models reported already in the literature is
the transition cumulative atomic multipole moments (TrCAMM) model
for estimation of the excitation energy transfer (EET) couplings in the F{\"o}rster
limit\cite{Blasiak.Maj.Cho.Gora.JCTC.2015}, 
or the vibrational solvatochromic distributed multipole moments
(SolCAMM) for the determination of the vibrational frequency shifts
of spatially localized IR probes.\cite{Blasiak.Lee.Cho.JCP.2013}
In general, having the OED and $\lambda$, the effective cumulative atomic charges
and dipole moments are given by
\begin{subequations} \label{e:a1-v.oep-camm}
 \begin{align} \label{e:a1-v.oep-camm.q}
  q_x^{\rm eff} &= \lambda Z_x - 
       \sum_{\alpha\in x}\sum_{\beta} 
        P_{\alpha\beta}^{\rm eff} \BraKet{\alpha}{\beta} \;, \\
  {\BM\upmu}_x^{\rm eff} &= 
       \sum_{\alpha\in x}\sum_{\beta} 
        P_{\alpha\beta}^{\rm eff}
   \left\{ 
    \BraKet{\alpha}{\beta} {\bf R}_x - \tBraKet{\alpha}{\hat{\bf r}}{\beta} 
   \right\}
   \;,
 \end{align}
\end{subequations}
where ${\bf R}_x$ is the position vector of the $x$-th nucleus.
Note that $P_{\alpha\beta}^{\rm eff}$ can refer to the particular electron
spin, or their sum as a bond order matrix as well.
Equations for the effective distributed quadrupoles, octupoles and hexadecapoles
can be easily obtained by analogy from Ref.~\cite{Blasiak.Maj.Cho.Gora.JCTC.2015}

% -----------------------
\bibliography{references}
% -----------------------

\end{document}

% numbering for Figures in SI
\renewcommand{\thefigure}{S\arabic{figure}}
\setcounter{figure}{0}

\title{Supplementary Information\\Ab Initio Effective One-Electron Potential Operators. I.
Applications for Charge-Transfer Energy in Effective Fragment Potentials}

\author{Bartosz B{\l}asiak}
\email[]{blasiak.bartosz@gmail.com}
\homepage[]{https://www.polonez.pwr.edu.pl}

\author{Joanna D. Bednarska}
\author{Marta Cho{\l}uj} 
\author{Wojciech Bartkowiak}

\affiliation{Department of Physical and Quantum Chemistry, Faculty of Chemistry, 
Wroc{\l}aw University of Science and Technology, 
Wybrze{\.z}e Wyspia{\'n}skiego 27, Wroc{\l}aw 50-370, Poland}

\date{\today}
\maketitle
\tableofcontents

\section{Asymptotic dependence of CT interactions for methanol-water complex}
\begin{figure}[ht]
\includegraphics[width=0.5\textwidth]{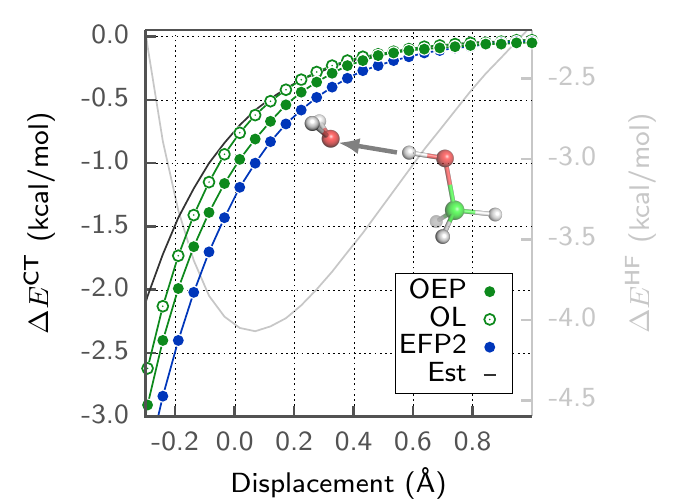}
\caption{\label{f:fig-s1} {\bf Asymptotic dependence of the charge transfer energy
in the CT/HF0 formulation for water-methanol complex.} 
Water molecule has been translated by $R$ from the starting geometry
along the vector specified in the inset picture.
The counterpoise\hyp{}corrected total interaction energy
is also shown for comparison in light grey color in this figure.
All data were obtained at HF/6-311++G(d,p) level of theory
with the EDF-1 scheme and aug-cc-pVQZ-jkfit auxiliary basis set.
} 
\end{figure}
\clearpage

\section{Validation of EFP2 implementation vs GAMESS US}

\begin{figure}[ht]
\includegraphics[width=0.8\textwidth]{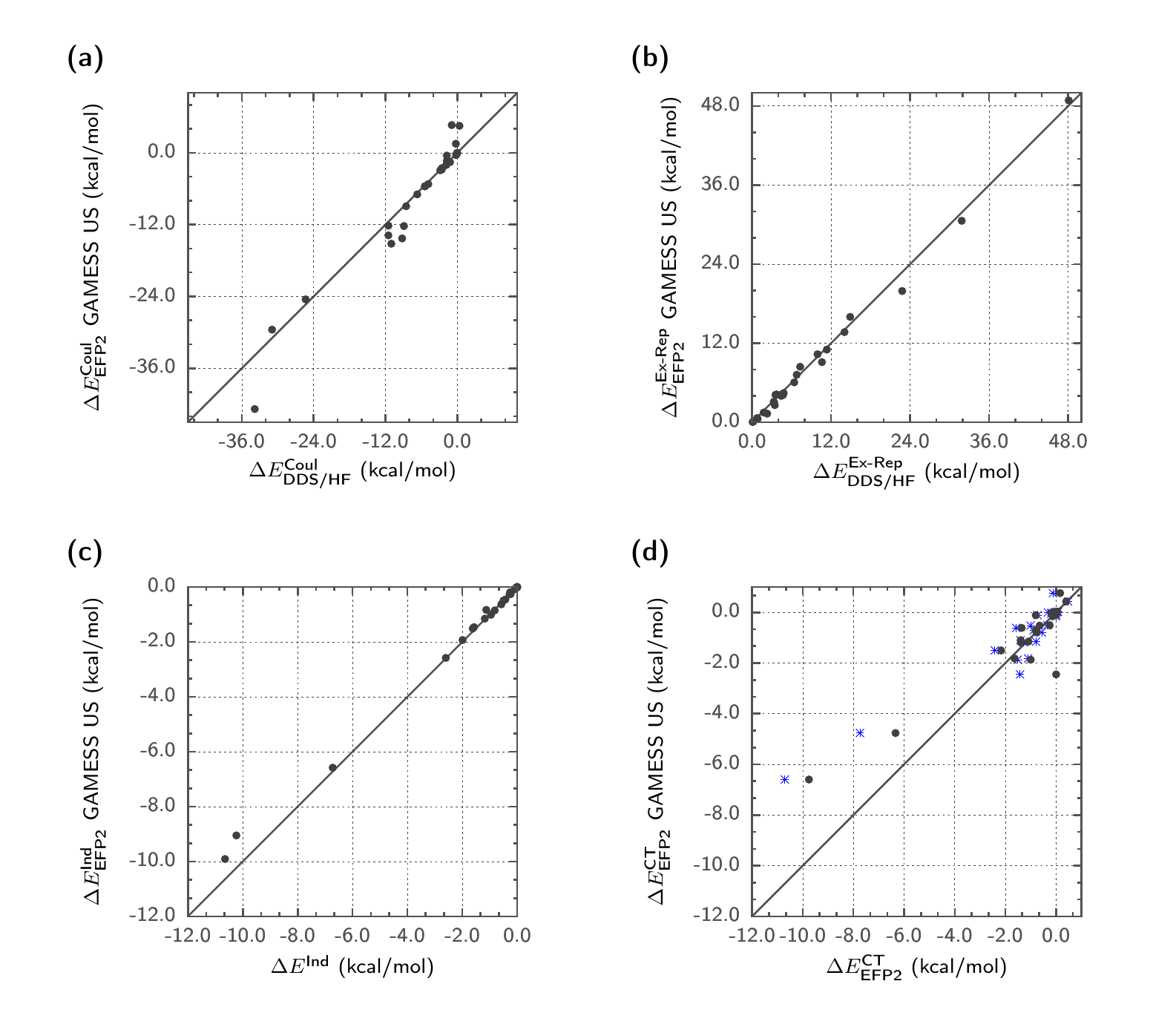}
\caption{\label{f:fig-s2} {\bf Validation of implementation of the interaction energy components
with respect to the EFP2 GAMESS US code.} 
Results obtained for the NCB31 database set assuming 6-311++G(d,p) primary basis set. 
(a) - Coulombic, (b) - exchange-repulsion, (c) - induction, and (d) - charge-transfer
interaction energy. In (d), black circles and blue crosses correspond to the
CT EFP2 energies with CAMM up to distributed quadrupoles and octupoles, respectively.
For more details see the main text (Section III, \emph{Calculation Details}).
} 
\end{figure}

In Figure~\ref{f:fig-s2} interaction energy components
obtained by using the code developed in this work are
compared to the ones obtained by using the EFP2 calculation routines
in GAMESS US quantum chemistry package\cite{GAMESS.JCC.1993} (version: Sept 30, 2017 R2 Public Release). 
For this, EFP2 parameters
were generated by running the \texttt{MAKEFP} routine
with \texttt{CTVVO} option switched to \texttt{.FALSE.} which corresponds
to using all the canonical molecular orbitals from HF calculations
instead of valence virtual orbitals.
Generally, Coulombic energies (Figure~S2(a)) 
are in very good agreement with a few exeptions in the PPS5 set,
for which the differences are due to the inaccuracies in the DMTP
expansion used in the EFP2 model -- note that the Coulombic component in the DDS/HF method
is free from multipole approximation and is equivalent to the first\hyp{}order SAPT Coulombic energy.
Exchange\hyp{}repulsion and induction energies (Figures~S2(b) and S2(c), respectively)
are in very good agreement. CT energies (Figure~S2(d)) 
evaluated with CAMM up to quadrupoles for the potential energy matrix elements (black filled circles)
are generally also in decent quantitative agreement,
except for NH$_3$--FCl, (HCOOH)$_2$ and (HCONH$_2$)$_2$ 
%and parallel\hyp{}displaced benzene dimer
where the differences
are between 2--4 kcal/mol. Only in the case of the latter three systems,
adding distributed octupoles (blue crosses in Figure~S2(d)) 
lowers the CT energy values by around 1 kcal/mol,
with negligible changes for the rest.
\clearpage

\section{Auxiliary basis set dependence of CT/HF0 energy}
\begin{figure}[ht]
\includegraphics[width=0.5\textwidth]{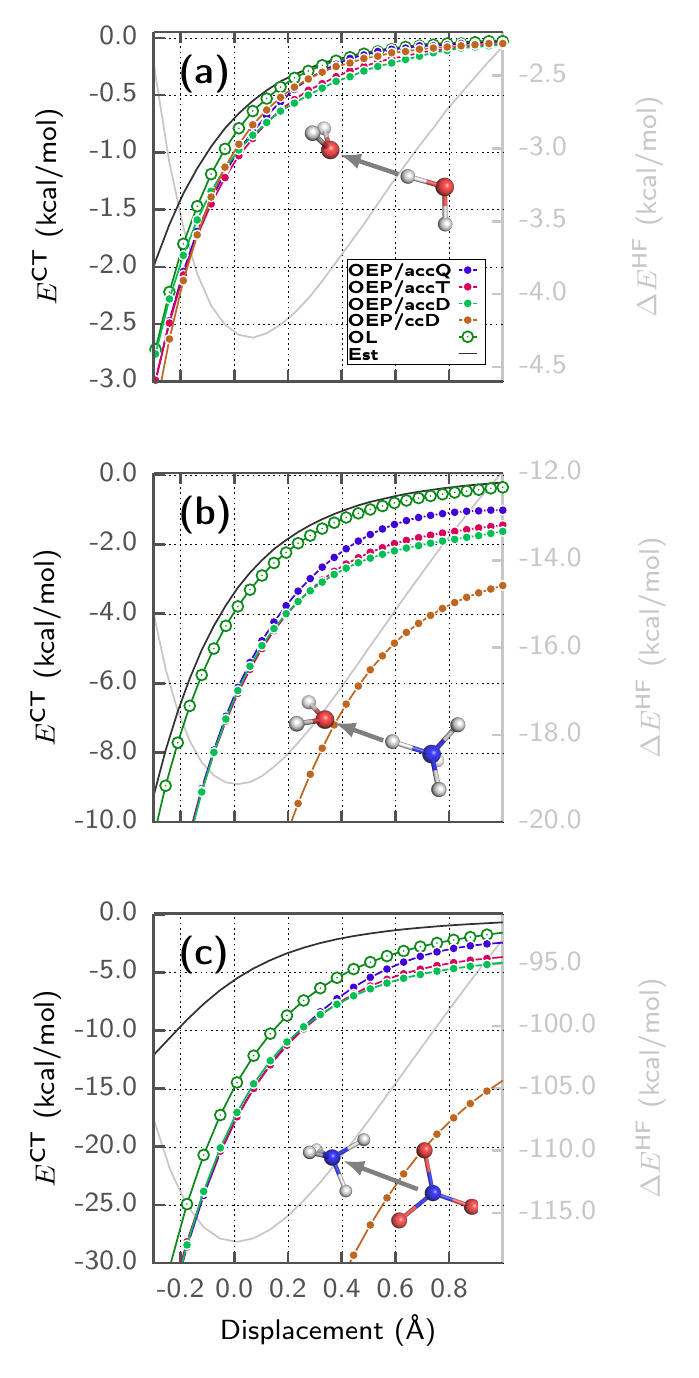}
\caption{\label{f:fig-s3} {\bf Effect of auxiliary basis set size
on the asymptotic dependence of CT/HF0 energies in model bi\hyp{}molecular complexes.} 
Auxiliary basis sets were abbreviated for convenience as follows:
ccD -- cc-pVDZ, accX -- aug-cc-pVXZ where X=D,T,Q.
} 
\end{figure}
\clearpage

% -----------------------
\bibliography{references}
% -----------------------

%% file: header.tex
\usepackage{graphicx}% Include figure files
\usepackage{dcolumn}% Align table columns on decimal point
\usepackage{bm}% bold math

\usepackage{hyphenat}

\usepackage{multirow}
\usepackage{booktabs}

\usepackage{amsmath}
\usepackage{amsfonts}
\usepackage{amssymb}
\usepackage{amsbsy}
\usepackage{mathrsfs}
\usepackage{upgreek}
\usepackage[cbgreek]{textgreek}

\newcommand{\Bra}[1]{\ensuremath{\bigl\langle {#1} \bigl\lvert}}
\newcommand{\Ket}[1]{\ensuremath{\bigr\rvert {#1} \bigr\rangle}}
\newcommand{\BraKet}[2]{\ensuremath{\bigl\langle {#1} \bigl\lvert {#2} \bigr\rangle}}
\newcommand{\tBraKet}[3]{\ensuremath{\bigl\langle {#1} \bigl\lvert {#2} \bigl\lvert {#3} \bigr\rangle}}
\newcommand{\BM}[1]{\bm{#1}}
\graphicspath{{./figures/}}